\documentclass{iopart}

\usepackage[dvips]{graphics}
\usepackage{graphicx}
\usepackage{psfrag}

\renewcommand{\u}{\underline}
\renewcommand{\b}{\mathbf}
\renewcommand{\o}{\overline}

\begin{document}

\title{Investigation of the nonlocal coherent-potential approximation}

\author{D.~A.~Rowlands}
\address{H.H.~Wills Physics Laboratory, University of Bristol, Bristol BS8 1TL, U.K.}

\begin{abstract}

Recently the nonlocal coherent-potential approximation (NLCPA) has been introduced by Jarrell and Krishnamurthy for describing the electronic
structure of substitutionally-disordered systems. The NLCPA provides systematic corrections to the widely used coherent-potential
approximation (CPA) whilst preserving the full symmetry of the underlying lattice. Here an analytical and systematic numerical study of the
NLCPA is presented for a one-dimensional tight-binding model Hamiltonian, and comparisons with the embedded cluster method (ECM) and
molecular coherent potential approximation (MCPA) are made.

\end{abstract}


\submitto{\JPCM}

\section{Introduction}

For many years the coherent-potential approximation (CPA) \cite{Soven1} has been widely used for describing the electronic structure of
substitutionally-disordered systems. However as a single-site mean-field theory, the CPA is not a fully satisfactory theory of disorder and it
leaves much important physics out of consideration. In the CPA a site only feels the average effect of its environment and so, for example,
statistical fluctuations in the chemical environment of a site responsible for sharp structure in the density of states (DOS) are not
described \cite{Gonis1}. Furthermore the CPA is incapable of describing the effects of chemical short-range order (SRO) and other important
environmental effects such as lattice displacements. However a new theory, the nonlocal coherent-potential approximation (NLCPA)
\cite{Jarrell1}, has recently been introduced by Jarrell and Krishnamurthy as a way of addressing such problems. The purpose of this paper is
to investigate the general validity of the NLCPA by means of numerical calculations with a simple one-dimensional (1D) tight-binding model.

There have been many previous attempts to generalize the CPA, though none have been completely satisfactory \cite{Gonis1}. Of these theories,
only the embedded cluster method (ECM) \cite{Gonis6} and the molecular coherent-potential approximation (MCPA) \cite{Tsukada1} will be
considered here. The ECM is a method for embedding a cluster of real impurity sites into a medium such at that provided by the CPA. Averaging
over an ensemble of the various cluster configurations can therefore give a non self-consistent description of local environment effects. The
MCPA goes a step further by enforcing self-consistency with respect to the cluster, and so yields a new effective medium. However this medium
has the periodicity of the cluster and so possesses the unsatisfactory property of breaking the symmetry of the underlying lattice. Such
supercell periodicity also means the MCPA is computationally prohibitive since it involves a Brillouin zone (BZ) integration which scales
with the cluster size, thus preventing its application to realistic systems.

However the recently proposed NLCPA \cite{Jarrell1} appears to possess all the attributes required for a satisfactory cluster generalization
of the CPA. It was introduced in the context of a two-dimensional tight-binding model Hamiltonian as the static version of the dynamical
cluster approximation (DCA) \cite{Hettler1,Hettler2} used for describing short-range correlations in strongly-correlated electron systems. A
recent application has been to describe impurity bound states in disordered d-wave superconductors \cite{Moradian2}. The NLCPA again
determines an effective medium via the self-consistent embedding of a cluster, however the problem of maintaining translational invariance is
solved by imposing Born-von Karman boundary conditions on the cluster, leading to an effective medium which has the site-to-site
translational invariance of the underlying lattice. A further consequence is that the computational difficulties of the MCPA are alleviated
since the BZ integration in the NLCPA does not scale as the cluster size increases. 

The NLCPA was subsequently derived by Rowlands \cite{Rowlands1} within the first-principles Korringa-Kohn-Rostoker (KKR) 
\cite{Korringa1,Kohn2} multiple scattering formalism, and the resulting KKR-NLCPA \cite{Rowlands1,Rowlands2} method was first implemented for
a realistic three-dimensional system ($bcc$ $Cu_{50}Zn_{50}$) in \cite{Rowlands3}. Moreover, the KKR-NLCPA method has recently been combined 
with density functional theory, enabling first-principles total energy calculations to be carried out as a function of SRO using 
self-consistently determined potentials \cite{Rowlands5}. Such self-consistent-field (SCF)-KKR-NLCPA calculations \cite{Rowlands5} naturally 
include a description of the Madelung contribution to the total energy which is missing in the conventional single-site SCF-KKR-CPA, and 
important future developments include application to systems with spin \cite{Staunton2}, strain \cite{Gyorffy9}, and valency \cite{Luders1} 
fluctuations.

Due to the continued conceptual development of the NLCPA method and its first-principles application to realistic systems in particular, it
is important to establish its general validity by addressing the following questions. Firstly, does the NLCPA produce physically meaningful
results, in other words are the fluctuations described real? Secondly, does the DOS calculated using the NLCPA converge to the exact result
in practice as the cluster size increases, and does it do so systematically? To date, these important questions have not been addressed.
However this can be done by carrying out a systematic study of the NLCPA for a simple 1D tight-binding model with diagonal disorder and
nearest neighbour hopping. There are three reasons for this. The first is that in 1D the exact result can be obtained against which
meaningful comparisons can be made. Secondly, fluctuations are much more significant in 1D and so detailed structure is expected in the DOS
which can be accurately interpreted. Thirdly, such a model is computationally very simple and so large clusters can be considered, thus
enabling the convergence properties of the theory to be investigated. Such an analysis is carried out in this paper, and any insight gained
which could aid the interpretation of realistic calculations is presented.

The outline of this paper is as follows. For completeness and for the purpose of introducing notation, the concept of the effective medium is
described first in section~\ref{definitions}. Next, brief derivations of the CPA, ECM, and MCPA are given in sections~\ref{cpa}, \ref{ecm},
and \ref{mcpa} respectively. In particular, the equivalence of the `cavity' \cite{Jarrell1,Moradian2} and the `renormalized interactor'
\cite{Gonis1} formalism is established. An understanding of these theories is crucial in order to appreciate and assess the advantages and
possible limitations of the NLCPA. In section \ref{nlcpa}, a derivation of the NLCPA is given with an attempt to provide a physical picture
of the theory, and comparisons with the formalism of sections~\ref{cpa}, \ref{ecm}, and \ref{mcpa} are made where necessary. In
section~\ref{graphs}, a numerical investigation of the NLCPA is presented, and results are compared with those of the CPA, ECM, and MCPA.
Finally, conclusions are drawn in section~\ref{conclude}.

\section{Formalism}\label{definitions}

In a basis labelled by the sites of the direct lattice, consider a tight-binding Hamiltonian with matrix elements
\begin{equation}\label{hbasis}
 H^{ij}=\epsilon^{i}\delta_{ij}+W^{ij}(1-\delta_{ij})
\end{equation}
and corresponding Dyson equation
\begin{equation}\label{propagator}
 G^{ij}=G^{ij}_{0}+\sum_{k}G^{ik}_{0}\epsilon^{k}G^{kj}
\end{equation} 
where $\epsilon^{i}$ is the on-site energy at site $i$, and $W^{ij}$ is the parameter describing the hopping between site $i$ and site $j$. 
For the case of a random substitutionally-disordered binary alloy with only diagonal disorder, the set of $\{\epsilon^i\}$ vary from site to
site in a random fashion, taking the value $e_A$ with probability $c$ or $e_B$ with probability $1-c$. Denoting a specific configuration of
$\{\epsilon^i\}$ by $\gamma$, and the corresponding Green's function for that configuration by $G_{\gamma}$, we are interested in
$\left<G_{\gamma}\right>$, the average over all possible configurations. To do that we iterate (\ref{propagator}) in a Born series, average
term by term, and resum in the form
\begin{equation}\label{dyson}
 \left<G^{ij}_{\gamma}\right>=G^{ij}_{0}+\sum_{k,l}G^{ik}_{0}\Sigma^{kl}\left<G^{lj}_{\gamma}\right>
\end{equation}
where $\Sigma^{ij}$ is the exact \emph{self-energy}.~\footnote{Clearly $\Sigma^{ij}$ is not diagonal in the site index otherwise  we would be
making a mean-field approximation by writing the average of a product $\left<\epsilon^{i}G^{ij}_{\gamma}\right>$ in terms of the product of
the averages $\left<\epsilon^{i}\right>\left<G^{ij}_{\gamma}\right>$.} Since (\ref{dyson}) is also a Dyson equation, it may be interpreted as
describing an \emph{effective medium} in which the site-diagonal part of $\Sigma^{ij}$ decribes an effective on-site energy, and the site
off-diagonal part decribes an effective correction to the hopping. The quantities $\left<G^{ij}_{\gamma}\right>$ and $\Sigma^{ij}$ both
possess the full translational (and point-group) symmetry of the underlying lattice, and so (\ref{dyson}) may be expressed in reciprocal 
space as
\begin{equation}\label{dysonk}
 \left<G(\b{k})\right>=G_0(\b{k})+G_0(\b{k})\Sigma(\b{k})\left<G(\b{k})\right>
\end{equation}
The matrix elements of the Green's function are therefore given by the BZ integral
\begin{equation}\label{dysonij}
 \left<G^{ij}_{\gamma}\right>=\frac{1}{\Omega_{BZ}}\int_{\Omega_{BZ}}d\b{k}\,\left((G_0(\b{k}))^{-1}-\Sigma(\b{k})\right)^{-1}
                                e^{i\b{k}(\b{R}_i-\b{R}_j)}
\end{equation}
with $(G_0(\b{k}))^{-1}=E-W(\b{k})$. Since it is not feasible to solve the problem exactly, the idea of an effective medium theory is to 
determine the best possible approximation to $\Sigma^{ij}$ and $\left<G^{ij}_{\gamma}\right>$.

\subsection{Coherent-potential approximation}\label{cpa}

The main approximation made by the CPA \cite{Soven1} is to assume a site-diagonal translationally-invariant self-energy 
$\o{\epsilon}^{i}\delta_{ij}$, where $\o{\epsilon}^i=\o{\epsilon}$ for all $i$. The CPA effective medium is then described by the equation
\begin{equation}\label{dysoncpa}
  \o{G}^{ij}=G^{ij}_{0}+\sum_{k}G^{ik}_{0}{\o{\epsilon}}^{k}\o{G}^{kj}
\end{equation}
In order to determine the medium, let us consider any site $i$. By removing the sum over all sites $k$ and making up for the neglected terms
by replacing the free particle Green's function with the cavity Green's function ${\cal{G}}^{ii}$, the site-diagonal part of (\ref{dysoncpa})
at site $i$ can be formally rewritten in the form
\begin{equation}\label{dysoncavity}
\o{G}^{ii}={\cal{G}}^{ii}+{\cal{G}}^{ii}{\o{\epsilon}}^{i}\o{G}^{ii}
          =\left(\left({\cal{G}}^{ii}\right)^{-1}-\o{\epsilon}^{i}\right)^{-1}
\end{equation}
In fact ${\cal{G}}^{ii}$ is related to a quantity known as the renormalized interactor $\Delta^{ii}$ \cite{Gonis1} by
\begin{equation}\label{cavdelta}
   {\cal{G}}^{ii}=(E-\Delta^{ii})^{-1}
\end{equation}
where $\Delta^{ii}$ is given by the locator expansion \cite{Gonis1}
\numparts
\begin{eqnarray}
	\Delta^{ii}&=&\sum_{k{\neq}i}W^{ik}g^{k}W^{ki}+\sum_{k{\neq}i,l{\neq}i}W^{ik}g^{k}W^{kl}g^{l}W^{li}+\cdots\label{delta}\\
	 g^{i}&=&(E-\o{\epsilon}^i)^{-1}\label{gi}	 
\end{eqnarray}
\endnumparts
It can be seen that $\Delta^{ii}$ and therefore ${\cal{G}}^{ii}$ depends only on the medium surrounding site $i$ and is independent of the
chemical occupation of $i$ itself. It is therefore straightforward to define the Green's function for a real \emph{impurity} embedded 
in the medium simply by replacing the effective site energy $\o{\epsilon}^i$ with a real site energy $\epsilon_{\alpha}^{i}$ at site $i$,
where $\alpha=$ $A$ or $B$. From (\ref{dysoncavity}) this is given by
\begin{equation}
	G^{ii}_{\alpha}=\left(\left({\cal{G}}^{ii}\right)^{-1}-\epsilon^{i}_{\alpha}\right)^{-1}								 
	               =\left(\o{\epsilon}^{i}-\epsilon^{i}_{\alpha}+(\o{G}^{ii})^{-1}\right)^{-1}
\end{equation}
Now the CPA demands that
\begin{equation}\label{sccpa}
	\sum_{\alpha}P_{\alpha}G^{ii}_{\alpha}=\o{G}^{ii}
\end{equation}
where $P_{\alpha}$ is the probability that site $i$ is of chemical type $\alpha$. In other words, the replacement of an effective site energy
by a real site energy should, on the average, produce no change to the CPA medium. Since the medium is translationally-invariant, it follows
from (\ref{dysoncpa}) that $\o{G}^{ii}$ must also be given by the BZ integral
\begin{equation}\label{sccpa2}
 \o{G}^{ii}=\frac{1}{\Omega_{BZ}}\int_{\Omega_{BZ}}d\b{k}\,\left(E-\o{\epsilon}-W(\b{k})\right)^{-1}
\end{equation}
The CPA medium is therefore determined from a self-consistent solution of (\ref{sccpa},\ref{sccpa2}).

\subsection{Embedded cluster method}\label{ecm}

The ECM \cite{Gonis6} is a method for embedding a cluster of real impurity sites in a medium such as that provided by the CPA. First consider
(\ref{dysoncpa}) describing the CPA medium, but with the sites $i,j$ lying within some cluster $C$. By restricting the sum over all
sites $k$ to run over the sites belonging to the cluster only, and making up for the neglected terms by replacing the free particle Green's
function with the cavity Green's function ${\cal{G}}$, (\ref{dysoncpa}) can be formally rewritten in the form
\begin{equation}\label{dysonecmIJ}
	\o{G}^{IJ}={\cal{G}}^{IJ}+\sum_{K}{\cal{G}}^{IK}\o{\epsilon}^{K}\o{G}^{KJ}
\end{equation}
where notation has been introduced so that all sites belonging to the cluster have been denoted by capital letters. In matrix form
(\ref{dysonecmIJ}) can be expressed as
\begin{equation}\label{dysonecmC}
\o{\u{G}}^{CC}={\cal{\u{G}}}^{CC}+{\cal{\u{G}}}^{CC}\o{\u{\epsilon}}^{C}\o{\u{G}}^{CC}
              =\left(({\cal{\u{G}}}^{CC})^{-1}-\o{\u{\epsilon}}^{C}\right)^{-1}
\end{equation}
where all matrices are in the space of the sites belonging to the cluster. In analogy to (\ref{cavdelta}) in the derivation of the
CPA, ${\cal{\u{G}}}^{CC}$ is related to the \emph{cluster} renormalized interactor $\u{\Delta}^{CC}$ \cite{Gonis1} by the 
expression
\begin{equation}
   {\cal{\u{G}}}^{CC}=(\u{E}-\u{\Delta}^{CC})^{-1}
\end{equation}
where $\Delta^{IJ}$, the matrix elements of $\u{\Delta}^{CC}$, are given by the expansion
\begin{equation}
\Delta^{IJ}=\sum_{k\not{\in}C}W^{Ik}g^{k}W^{kJ}+\sum_{k\not{\in}C,l\not{\in}C}W^{Ik}g^{k}W^{kl}g^{l}W^{lJ}+\cdots 
\end{equation}
and $g^i$ is given by (\ref{gi}). It can be seen that $\u{\Delta}^{CC}$ and therefore ${\cal{\u{G}}}^{CC}$ depends only on the
medium surrounding the cluster $C$ and is independent of the chemical occupation of $C$ itself. The Green's function for an \emph{impurity 
cluster} embedded in the CPA medium can now be straightforwardly defined simply by replacing the effective CPA cluster-site energy matrix 
$\o{\u{\epsilon}}^C$ with a matrix $\u{\epsilon}^{C}_{\gamma}$ containing a configuration $\gamma$ of real site energies 
$\{\epsilon_{\alpha}^{I}\}$ at each cluster site $I$, where $\alpha=$ $A$ or $B$. From (\ref{dysonecmC}) this is given by
\begin{equation}\label{ecmimpC}
	\u{G}^{CC}_{\gamma}=\left(\left({\cal{\u{G}}}^{CC}\right)^{-1}-\u{\epsilon}^{C}_{\gamma}\right)^{-1}
	                   =\left(\o{\u{\epsilon}}^{C}-\u{\epsilon}^{C}_{\gamma}+(\o{\u{G}}^{CC})^{-1}\right)^{-1}
\end{equation}
Once the CPA medium is known, the Green's function for an impurity cluster embedded in the medium can be straightforwardly determined through
(\ref{ecmimpC}), and properties can then be calculated at any desired cluster site. However an average over all possible cluster
configurations $\gamma$ is usually taken, in which case it will be found that $\left<\u{G}^{CC}_{\gamma}\right>\neq{\o{\u{G}}^{CC}}$ for a
cluster size $N_c>1$. This is due to an average effect on the electron hopping within the cluster caused by the disorder configurations. Thus
$\o{\u{\epsilon}}^{C}$ gains an off-diagonal correction and could be interpreted as a (non-self-consistently determined) \emph{cluster
self-energy} $\u{\Sigma}^{C}$ with matrix elements $\Sigma^{IJ}$. However this is defined only at the cluster sites, with the single-site CPA
self-energy $\o{\epsilon}^{i}$ situated at all other sites, and hence there is no new effective medium. To do that, self-consistency with
respect to the cluster must be enforced. One way forward is the MCPA \cite{Tsukada1}, as described in the next subsection.

\subsection{Molecular coherent-potential approximation}\label{mcpa}

As described above, a cluster self-energy may be defined by averaging over an ensemble of impurity cluster configurations embedded in a
medium. The idea of the MCPA \cite{Tsukada1} is to determine an effective medium with a periodically-repeating cluster self-energy. Such a
medium may be defined by first dividing the lattice up into a collection of identical non-overlapping clusters, and then writing the Green's
function in the form
\begin{equation}\label{dysonmcpa}
    \o{\u{G}}^{CC'}=\u{G}_{0}^{CC'}+\sum_{C''}\u{G}_{0}^{CC''}\left(\u{\Sigma}^{C''}+\u{W}^{C''C''}\right)\o{\u{G}}^{C''C'}
\end{equation}
These cluster matrices have site matrix elements
\begin{equation}
  \left[\u{\o{G}}^{CC'}\right]_{ij}=\o{G}^{ij} , \quad 
	\left[\u{G}_{0}^{CC'}\right]_{ij}=G_{0}^{ij} , \quad i\in C,\ j\in C' 
\end{equation}
\begin{equation}
		\left[\u{\Sigma}^{C}\delta_{CC'}\right]_{ij}=\Sigma^{ij} , \quad
		\left[\u{W}^{CC}\delta_{CC'}\right]_{ij}=W^{ij} , \quad i,j \in C 
\end{equation}
The main approximation made here has been to assume that the (as of yet undetermined) cluster self-energy is \emph{cluster-diagonal}. This
means there are self-energy terms relating sites in the same cluster, but none relating sites in different clusters. The single-site
translational invariance of the underlying lattice is therefore broken. However, the MCPA does require the medium to be invariant upon
translation by a cluster so that it has the periodicity of a `super-lattice'. As such, $\u{\Sigma}^{C}$ is independent of the cluster label
$C$, and the Green's function $\o{\u{G}}^{CC'}$ depends only on the difference between $C$ and $C'$. The real-space cluster dependence of
(\ref{dysonmcpa}) may therefore be removed using the Fourier transform
\begin{equation}\label{mcpaft}
    \o{\u{G}}^{CC}(\b{q})=\sum_{{C'}}\o{\u{G}}^{CC'}e^{-i\b{q}(\b{R}_{C}-\b{R}_{C'})}
\end{equation}
where $\b{R}_{C}-\b{R}_{C'}$ is the vector distance between the centres of clusters $C$ and $C'$, and $\b{q}$ is a vector in the (smaller) BZ
of the super-lattice. Equation~(\ref{dysonmcpa}) now becomes
\begin{equation}\label{dysonq}
    \o{\u{G}}^{CC}(\b{q})=\u{G}_{0}^{CC}(\b{q})+\u{G}_{0}^{CC}(\b{q})\left(\u{\Sigma}^{C}+\u{W}^{CC}\right)\o{\u{G}}^{CC}(\b{q})
\end{equation}
where all matrices have the dimension of the cluster size only, and the inter-cluster dependence is expressed through the reciprocal-space 
vector $\b{q}$. 

The next step is to embed an ensemble of impurity cluster configurations into the medium defined above. This may done using the ECM of 
section~\ref{ecm}, yielding the impurity cluster Green's function
\begin{equation}
	\u{G}^{CC}_{\gamma}=\left(\u{\Sigma}^{C}-\u{\epsilon}^{C}_{\gamma}+(\o{\u{G}}^{CC})^{-1}\right)^{-1}
\end{equation}
However, there are two main differences. Firstly, the cluster self-energy $\u{\Sigma}^C$ replaces the effective CPA cluster-site energy
matrix $\u{\epsilon}^C$. Secondly, the medium outside the cluster comprises of a periodically repeating cluster self-energy rather than
single-site CPA effective site energies, and hence $\u{\Delta}^{CC}$ has a slightly different expansion~\cite{Gonis1}. The final step is to
generalize the CPA argument and demand that the average of $\u{G}^{CC}_{\gamma}$ over all $\gamma$ be equal to the Green's function for the
medium itself i.e.
\begin{equation}\label{scmcpa}
    \left<\u{G}^{CC}_{\gamma}\right>=\o{\u{G}}^{CC}\quad{or}\quad\sum_{\gamma}P_{\gamma}G^{IJ}_{\gamma}=\o{G}^{IJ}\quad{\forall\,{I,J}\in{C}}
\end{equation}
For a cluster containing $N_c$ sites, the number of configurations will be $2^{N_c}$ for a binary alloy. Since the medium has the periodicity
of the cluster, it follows from (\ref{mcpaft},\ref{dysonq}) that $\o{\u{G}}^{CC}$ must also satisfy
\begin{eqnarray}\label{bzmcpa}
\o{\u{G}}^{CC}=\frac{1}{\Omega_{BZ}}\int_{\Omega_{BZ}}d\b{q}\left(\u{E}-\u{\Sigma}^{C}-\u{W}^{CC}-\u{W}^{CC}(\b{q})\right)^{-1}
\end{eqnarray}
where the integral is over the first BZ of the super-lattice. Therefore the MCPA medium is determined from a self-consistent
solution of (\ref{scmcpa},\ref{bzmcpa}).

The MCPA appears to be a natural generalization of the single-site CPA, but it does have some major disadvantages. Specifically, the
intra-cluster hopping is now different from the inter-cluster hopping expressed through $\u{W}^{CC}(\b{q})$ and so measuring properties at 
sites situated at the boundary of the cluster will in general give different results than sites at the centre of the cluster, for example. 
The supercell periodicity also causes problems in obtaining a well-defined spectral function, and leads to the computationally-demanding BZ
integration in (\ref{bzmcpa}) which scales as $N_c$ increases. 

\subsection{Nonlocal coherent-potential approximation}\label{nlcpa}

Like the MCPA, the NLCPA \cite{Jarrell1} determines an effective medium via the self-consistent embedding of a cluster. However, the key
difference is that \emph{Born-von Karman boundary conditions} are imposed on the cluster, leading to an effective medium which has the
translational invariance of the underlying lattice. The first step is to solve the problem of an isolated cluster with Born-von Karman
boundary conditions imposed. 

\subsubsection{Cluster with Born-von Karman boundary conditions}\label{bvk}

\begin{figure}[!]
 \begin{center}
  \psfrag{a}[B1][B1][1.5][0]{$a$}
  \psfrag{-pi/a}[B1][B1][1.5][0]{$\frac{-\pi}{a}$}
	\psfrag{pi/a}[B1][B1][1.5][0]{$\frac{\pi}{a}$}
 \scalebox{0.6}{\includegraphics{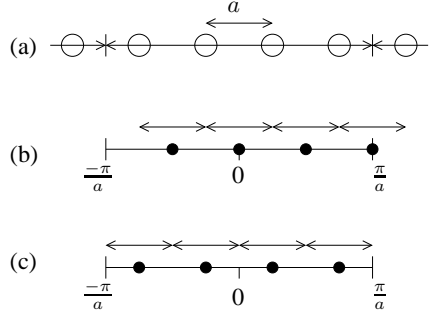}}   
 \caption{(a) Real space 1D tile (denoted by double-headed arrow of length $4a$) for a $N_c=4$ cluster. Sites are denoted by open circles, 
	and $a$ is the lattice constant. (b) Set of cluster momenta (periodic) denoted by closed circles for the $N_c=4$ cluster. The tiles centred
	at the cluster momenta are denoted by arrows and the solid line is the first BZ. The part of the tile centred at $\pi/a$ that lies outside
	the first BZ can be translated by reciprocal lattice vectors into the first BZ to lie between $-\pi/a$ and $-\pi/4a$. (c) Set of cluster 
	momenta (anti-periodic) denoted by closed circles for the $N_c=4$ cluster. Again the tiles centred at the cluster momenta are denoted by 
	arrows and the solid line is the first BZ.}\label{cluster}
 \end{center}
\end{figure}

Conventionally in solid state theory the problem of a lattice comprising of a large number of sites with Born-von Karman boundary conditions
is considered \cite{Ashcroft1}. In 1D, for example, this essentially means a very long chain of sites where one end of the chain maps round
to the other end. If the lattice constant is $a$, then the first BZ extends from $-\pi/a$ to $+\pi/a$ in reciprocal space, and the number of
$\b{k}$ points in the BZ is equal to the number of sites in the chain. It follows that if the length of the chain is now decreased to contain
only say a cluster of $N_c$ sites, then the number of $\b{k}$ points in the BZ will then equal $N_c$. These are referred to as the set of
\emph{cluster momenta} $\{\b{K}_n\}$, where $n=1,..,N_c$. However, the boundaries of the BZ remain the same since the lattice constant is
unchanged. The real-space cluster sites $\{I\}$ (denoted by capital letters) and the corresponding set of cluster momenta $\{\b{K}_n\}$
satisfy the relation \cite{Jarrell1} 
\begin{equation}\label{IJK}
	\frac{1}{N_c}\sum_{\b{K}_n}e^{i\b{K}_n(\b{R}_I-\b{R}_J)}=\delta_{IJ}
\end{equation}
for which the conventional lattice Fourier transform is recovered when $N_c\rightarrow\infty$. For example, in 1D the (periodic) solutions
are 
\begin{eqnarray}
	\{\b{R}_I\}=&(I-1)a& \ \ \ I=1,..,N_c \nonumber\\
	\{\b{K}_n^{P}\}=&\left(\frac{2n-N_c}{N_c}\right)\frac{\pi}{a}& \ \ \ n=1,..,N_c
\end{eqnarray}
for $N_c$ even. To date, only periodic solutions have been considered in the literature. However, for a given cluster size there is also a 
solution of (\ref{IJK}) corresponding to anti-periodic Born-von Karman boundary conditions (where in 1D, for example, the wavefunction at  
one end of the chain maps back to minus the value at the other end). In 1D, the anti-periodic set of cluster momenta are given by
\begin{equation}
	\{\b{K}_n^{AP}\}=\left(\frac{2n-N_c-1}{N_c}\right)\frac{\pi}{a} \ \ \ n=1,..,N_c
\end{equation}
A 1D example for $N_c=4$ is shown in figure~\ref{cluster}. Notice that $\{\b{K}_n^{P}\}$ always includes the origin in reciprocal space,
whilst $\{\b{K}_n^{AP}\}$ are shifted (by $-\pi/(N_c\,{a})$ in 1D) and lie symmetric about the origin. For the remainder of this derivation
it is however convenient to simply choose one of the set of solutions above. A discussion concerning this choice is given in 
section~\ref{solutions}. Cluster quantities that are translationally-invariant can be related in real and reciprocal space through 
(\ref{IJK}), for example for the cluster self-energy we have
\begin{eqnarray}\label{clusterselfenergy}
	\Sigma_{cl}^{IJ}=\frac{1}{N_c}\sum_{\b{K}_n}\Sigma_{cl}(\b{K}_n)e^{i\b{K}_n(\b{R}_I-\b{R}_J)} \nonumber\\
	\Sigma_{cl}(\b{K}_n)=\sum_{J}\Sigma_{cl}^{IJ}e^{-i\b{K}_n(\b{R}_I-\b{R}_J)}
\end{eqnarray}

\subsubsection{Mapping cluster to lattice}

\begin{figure}[!]
 \begin{center}
 \scalebox{0.6}{\includegraphics{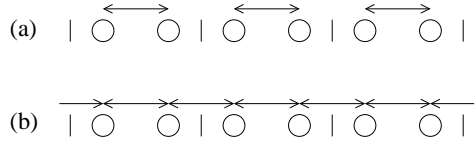}}   
 \caption{(a) Self-energy $\Sigma^{IJ}$ (denoted by double-headed arrows) for the case of a $N_c=2$ cluster in the MCPA, for $I\neq{J}$. The 
 vertical lines denote the boundaries of the tiles. (b) Same as (a) but for the NLCPA.}\label{selfenergy}
 \end{center}
\end{figure}

When mapping the cluster problem to the lattice, there are certain conditions on the geometry of the chosen cluster which must be satisified
in order that that the translational and point-group symmetry of the lattice is preserved \cite{Jarrell1}. Specifically, if the cluster sites
are now considered as `embedded' in the lattice, then it must be possible to surround them with a tile which has this point-group symmetry
and can be periodically repeated to fill out all space. This is analagous to the conventional Wigner-Seitz tile \cite{Ashcroft1} used to
surround a single site, and restricts the allowed values of $N_c$ for a given lattice. The construction for choosing appropriate clusters and
tiles was first explained in \cite{Jarrell1} for a 2D square lattice and was generalized to realistic 3D lattices in
\cite{Rowlands1,Rowlands2,Rowlands3}.~\footnote{This construction would also apply to the MCPA if aiming to preserve point-group symmetry.}
In 1D, the tiles are simply lines (see figure~\ref{cluster}).

Next, note that the periodicity of the real-space tiles define a reciprocal lattice, and each reciprocal lattice vector $\b{K}$ will
correspondingly be centred at a reciprocal-space tile. For $N_c=1$, this would correspond to the usual tiling of the conventional reciprocal
lattice with Brillouin zones. However, for $N_c>1$, reciprocal space will be more densely filled with $\b{K}$ points and the corresponding 
tiles will be smaller (by a factor of $N_c$). In the MCPA, such a tile would be the BZ of the `superlattice' (see section~\ref{mcpa}).
However, the idea of the NLCPA is to use $N_c$ such tiles to fill out the original BZ of the underlying lattice (see figure~\ref{cluster} for
a 1D example), thus preserving translational invariance. In fact, the relevant $\b{K}$ points will be the set of cluster momenta $\{\b{K}_n\}$.

The mapping of the cluster problem to the lattice may now be achieved as follows. In reciprocal space, consider the exact lattice self-energy
$\Sigma(\b{k})$ of (\ref{dysonk}). The first step is to average $\Sigma(\b{k})$ over the momenta $\b{q}$ within each of the $N_c$
tiles. This results in a `coarse-grained' lattice self-energy $\Sigma(\b{K}_n)$ which has a constant but different value within each tile. On
the other hand, the cluster self-energy $\Sigma_{cl}(\b{K}_n)$ is defined only at the cluster momenta $\{\b{K}_n\}$. The main approximation
made by the NLCPA is to set $\Sigma(\b{K}_n)$ to be equal to the value $\Sigma_{cl}(\b{K}_n)$ within each tile $n$ i.e.
\begin{equation}
	\frac{1}{\Omega_{\b{K}_n}}\int_{\Omega_{\b{K}_n}}d\b{q}\,\Sigma(\b{K}_n+\b{q})=\Sigma(\b{K}_n)\simeq\Sigma_{cl}(\b{K}_n)	
\end{equation}
In other words, the lattice self-energy is approximated by that obtained from the cluster. The NLCPA approximation to the lattice Green's
function in reciprocal space may now be represented by summing over the dispersion within each tile, yielding the set of coarse-grained 
values
\begin{equation}\label{GK}
	\o{G}(\b{K}_n)=\frac{N_c}{\Omega_{BZ}}\int_{\Omega_{\b{K}_n}}d\b{k}\left(E-W(\b{k})-\Sigma(\b{K}_n)\right)^{-1}
\end{equation}
which are straightforward to calculate since $\Sigma(\b{K}_n)$ is constant within each tile $\Omega_{\b{K}_n}$. Finally, using (\ref{IJK}), 
the real space Green's function at the cluster sites becomes
\begin{equation}\label{GIJK}
	\o{G}^{IJ}=\frac{1}{\Omega_{BZ}}\sum_{\b{K}_n}\int_{\Omega_{\b{K}_n}}
	            d\b{k}\left(E-W(\b{k})-\Sigma(\b{K}_n)\right)^{-1}e^{i\b{K}_n(\b{R}_I-\b{R}_J)}
\end{equation}
The reason for coarse-graining the Green's function is to preserve causality by removing phase factors within the cluster, however these can 
be reintroduced after the medium has been determined (see section~\ref{observables}).

A physical picture of the resulting NLCPA medium may be seen by examining the consequences of approximating the exact self-energy
$\Sigma(\b{k})$ by $\Sigma(\b{K}_n)$. Firstly, the self-energy is now restricted to act within the \emph{range}~\footnote{The self-energy
$\Sigma^{IJ}$ only takes into account exactly nonlocal correlations that are within the range of the cluster size. However, when the medium
is determined self-consistently, $\Sigma^{IJ}$ will include the effects of physics on a longer length scale at the mean-field level.} of the
cluster sites $\{I,J\}$ only (a more precise correlation length can be defined using Nyquist's sampling theorem \cite{Elliot1,Jarrell1}).
However, an electron can now propagate to any site in the lattice via the dispersion $W(\b{k})$, and will experience an identical self-energy
$\Sigma^{IJ}$ at that site. Thus $\Sigma^{IJ}$ remains a translationally-invariant quantity which depends only on the distance between sites
$I$ and $J$, now within the range of the cluster size, but independent of which site in the lattice is chosen to be site $I$. ($\Sigma^{IJ}$
will also possess the point-group symmetry of the underlying lattice provided that the clusters and tiles have been chosen correctly). An
illustration of this for the simplest case of a $N_c=2$ cluster in one dimension is shown in figure~\ref{selfenergy}.

\subsubsection{Impurity problem}

To determine the medium, the impurity problem must be solved. The first step is to define the reciprocal space cavity Green's function 
${\cal{G}}(\b{K}_n)$ via the Dyson equation
\begin{equation}\label{dysonKn}
    \o{G}(\b{K}_n)={\cal{G}}(\b{K}_n)+{\cal{G}}(\b{K}_n)\Sigma(\b{K}_n)\o{G}(\b{K}_n)
\end{equation}
In diagrammatic terms ${\cal{G}}(\b{K}_n)$ is introduced to `avoid over-counting self-energy diagrams on the cluster' \cite{Hettler1}.
Equation~(\ref{dysonKn}) can of course also be expressed in real space by applying the Fourier transform (\ref{IJK}) to yield
\begin{equation}\label{dysonreal}
    \o{G}^{IJ}={\cal{G}}^{IJ}+\sum_{K,L}{\cal{G}}^{IK}\Sigma^{KL}\o{G}^{LJ}
\end{equation}
where ${\cal{G}}^{IJ}$ is the real space cavity Green's function. In a similar manner to that introduced in the ECM and MCPA, here 
${\cal{G}}^{IJ}$ is independent of the chemical occupation of the cluster itself. However, the important difference is that it is not
possible to give an explicit real space expansion for ${\cal{G}}^{IJ}$ in terms of the cluster renormalized interactor as was possible in the
ECM and MCPA. The reason for this is due to the translational invariance of the self-energy in the NLCPA, and is an example of the so-called
`embedding problem' \cite{Gonis5}. If the ECM were used here to accomodate an impurity cluster in the NLCPA medium, then self-energy terms
linking the cluster sites to the medium sites would have to be broken (see figure~\ref{selfenergy}). Such a situation does not occur in the
MCPA since the self-energy is cluster diagonal and hence there are no self-energy terms linking sites in a cluster to the rest of the medium
in that case. Although the NLCPA cavity Green's function ${\cal{G}}^{IJ}$ is a mathematically well-defined quantity, it is not the same
quantity as that introduced in the ECM. The embedding of a cluster with Born-von Karman boundary conditions in the NLCPA medium should
therefore be viewed as a mathematical construction rather than a true embedding in the conventional sense. Nevertheless, the NLCPA impurity
cluster Green's function may be defined in analogy to (\ref{ecmimpC}) by replacing the cluster self-energy in (\ref{dysonreal}) with a 
particular configuration of site energies $\{\epsilon^I_{\alpha}\}$ i.e.
\begin{equation}
    G^{IJ}_{\gamma}={\cal{G}}^{IJ}+\sum_{K}{\cal{G}}^{IK}\epsilon_{\alpha}^{K}G^{KJ}_{\gamma}
\end{equation}
The NLCPA self-consistency condition is equivalent to that of the MCPA given in (\ref{scmcpa}) i.e.
\begin{equation}\label{scnlcpa}
    \sum_{\gamma}P_{\gamma}G^{IJ}_{\gamma}=\o{G}^{IJ}
\end{equation}
where $P_{\gamma}$ is the probability of configuration $\gamma$ occuring. The effective medium is therefore determined from a self-consistent
solution of (\ref{GIJK},\ref{scnlcpa}). The cluster probabilities depend on the SRO parameter $\alpha$ and so SRO may be included by
appropriately weighting the configurations in (\ref{scnlcpa}), provided that translational-invariance is preserved. Finally note that the
NLCPA formalism reduces to the CPA for $N_c=1$ and becomes exact as $N_c\rightarrow\infty$ since this would amount to solving the exact
problem described by (\ref{dyson},\ref{dysonk}).

\subsubsection{Calculation of Observables}\label{observables}

Once the medium has been determined through (\ref{GIJK},\ref{scnlcpa}), there is no longer any need to coarse-grain the Green's 
function via (\ref{GK}). Now, the Green's function may be calculated at any point in the BZ through
\begin{equation}\label{gkdisc}
	\o{G}(\b{k})=\left(E-W(\b{k})-\Sigma(\b{K}_{n})\right)^{-1}
\end{equation}
and correspondingly at any sites $i,j$ in the lattice by
\begin{equation}\label{gijdisc}
  \o{G}^{ij}=\frac{1}{\Omega_{BZ}}\int_{\Omega_{BZ}}d\b{k}\left(E-W(\b{k})-\Sigma(\b{K}_n)\right)^{-1}e^{i\b{k}(\b{R}_i-\b{R}_j)}
\end{equation}
In (\ref{gkdisc},\ref{gijdisc}) above, $\Sigma(\b{K}_n)$ takes the appropriate value within each tile $n$. The configurationally-averaged DOS
per site is given by the usual expression
\begin{equation}
  \o{n}(E)=-\frac{1}{\pi}\,Im\,\,\o{G}^{II}
\end{equation}
where $\o{G}^{II}$ may be calculated from (\ref{GIJK}) or from (\ref{gijdisc}) above. However, when calculating site-off diagonal observables
such as the spectral function, notice that $\Sigma(\b{K}_n)$ taking the appropriate (constant) value within each tile in (\ref{gkdisc}) leads
to discontinuities in $\o{G}(\b{k})$ at the tile boundaries. In order to remedy this, Maier~\etal \cite{Maier3} proposed a scheme
(within the DCA) to interpolate the self-energy using maximum entropy, however this can lead to causality being violated when the
fluctuations are large. An alternative scheme for removing such discontinuities has been proposed by Batt~\etal \cite{Batt1} (see
section~\ref{solutions} below). 

\subsubsection{Periodic and anti-periodic solutions}\label{solutions}

It was mentioned in section~\ref{bvk} that for a given cluster size there are two possible sets of cluster momenta which may be used,
$\{\b{K}_n^{P}\}$ and $\{\b{K}_n^{AP}\}$, corresponding to periodic or anti-periodic Born-von Karman boundary conditions respectively. In the
literature to date, only periodic boundary conditions have been considered. The validity of calculations using only $\{\b{K}_n^{P}\}$ is
analyzed in section~\ref{graphs}. However, it has been suggested that one should perform separate calculations using $\{\b{K}_n^{P}\}$ and
$\{\b{K}_n^{AP}\}$, and then average over both results \cite{Jarrell2}. The philosophy behind this argument is that this has the effect of
increasing the number of tiles, leading to a better sampling of the BZ. For example, in 1D the number of tiles is effectively doubled, as can
be seen from figure~\ref{cluster}. The validity of this suggestion is also examined in section~\ref{graphs}. 

However, Batt~\etal \cite{Batt1} have proposed a more sophisticated scheme for combining both solutions which also removes the
discontinuities in $\o{G}(\b{k})$ mentioned in section~\ref{observables} above. Firstly, let us label the Green's function (given by
(\ref{gkdisc})) for the periodic and anti-periodic solutions by $\o{G}_{P}(\b{k})$ and $\o{G}_{AP}(\b{k})$ respectively. Next, observe that
by construction $\o{G}_{P}(\b{k})$ is a better approximation to the exact result in the region of reciprocal space close to each of the
points $\{\b{K}_n^{P}\}$ (in fact $\o{G}_{AP}(\b{k})$ has discontinuities at $\{\b{K}_n^{P}\}$). Similarly $\o{G}_{AP}(\b{k})$ is a better
approximation to the exact result close to each of the points $\{\b{K}_n^{AP}\}$. This suggests that, in brief, one should construct a new
Green's function which follows $\o{G}_{P}(\b{k})$ close to each of the points $\{\b{K}_n^{P}\}$, and follows $\o{G}_{AP}(\b{k})$ close to
each of the points $\{\b{K}_n^{AP}\}$, and interpolates in-between. Such a combined solution is guaranteed to be causal since it will always
lie between the two causal extremes $\o{G}_{P}(\b{k})$ and $\o{G}_{AP}(\b{k})$. Furthermore, the problem of discontinuities is automatically
removed since neither solution is followed where it has a discontinuity. Full details together with an implementation of this scheme are
given in \cite{Batt1}.

\section{Results}\label{graphs}

\begin{figure}[!]
 \begin{center}
 \begin{tabular}{cc}
 \psfrag{(a)}[B1][B1][2][0]{(a)}\scalebox{0.33}{\includegraphics{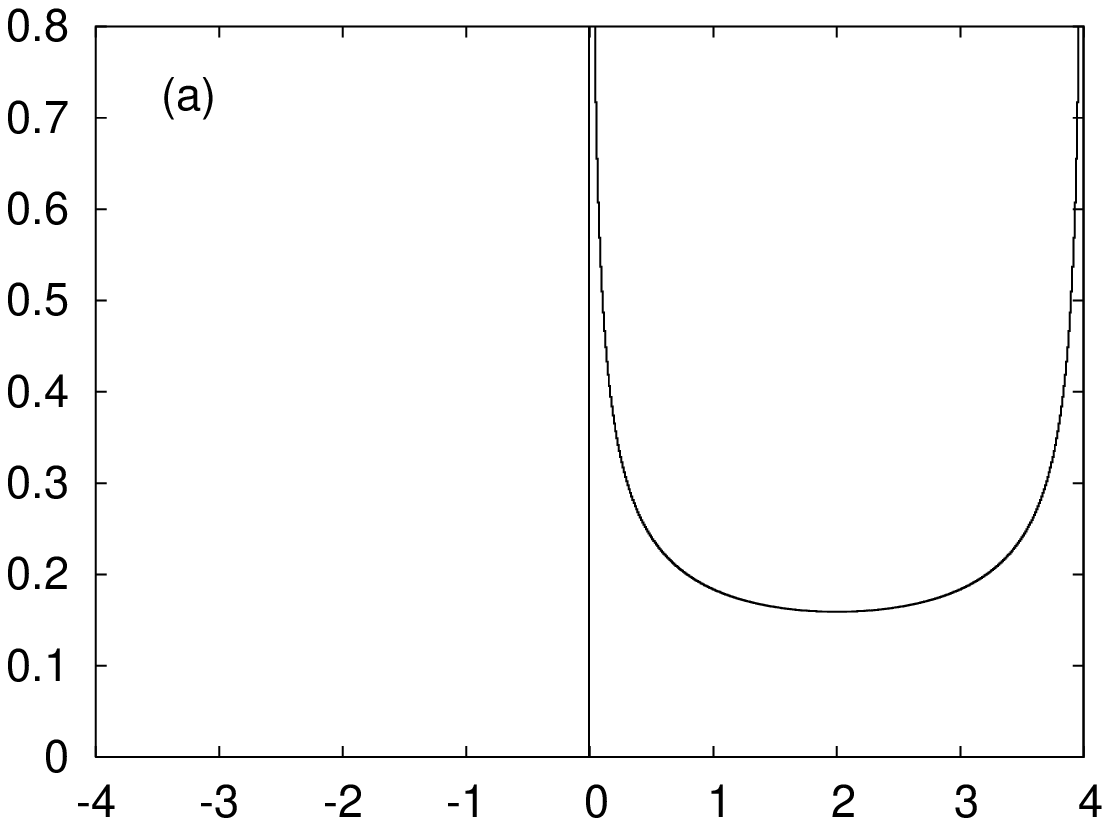}} &
 \psfrag{(c)}[B1][B1][2][0]{(c)}\scalebox{0.33}{\includegraphics{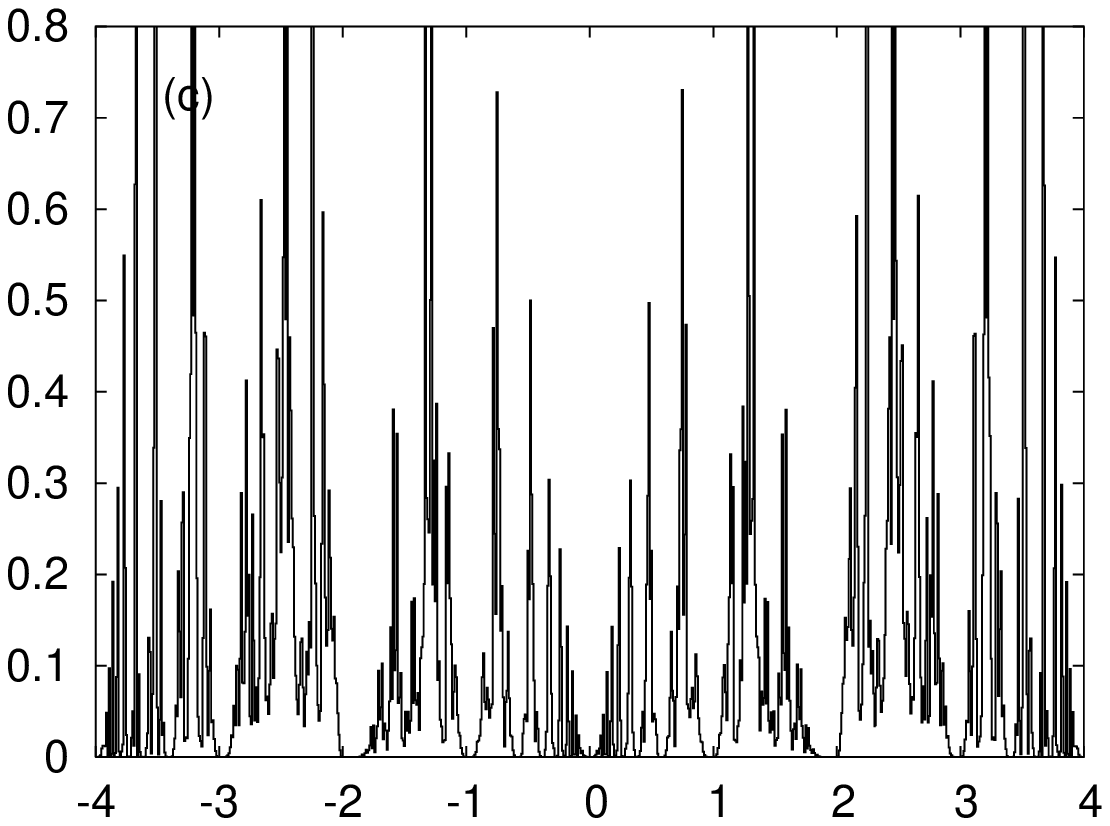}} \\
 \psfrag{(b)}[B1][B1][2][0]{(b)}\scalebox{0.33}{\includegraphics{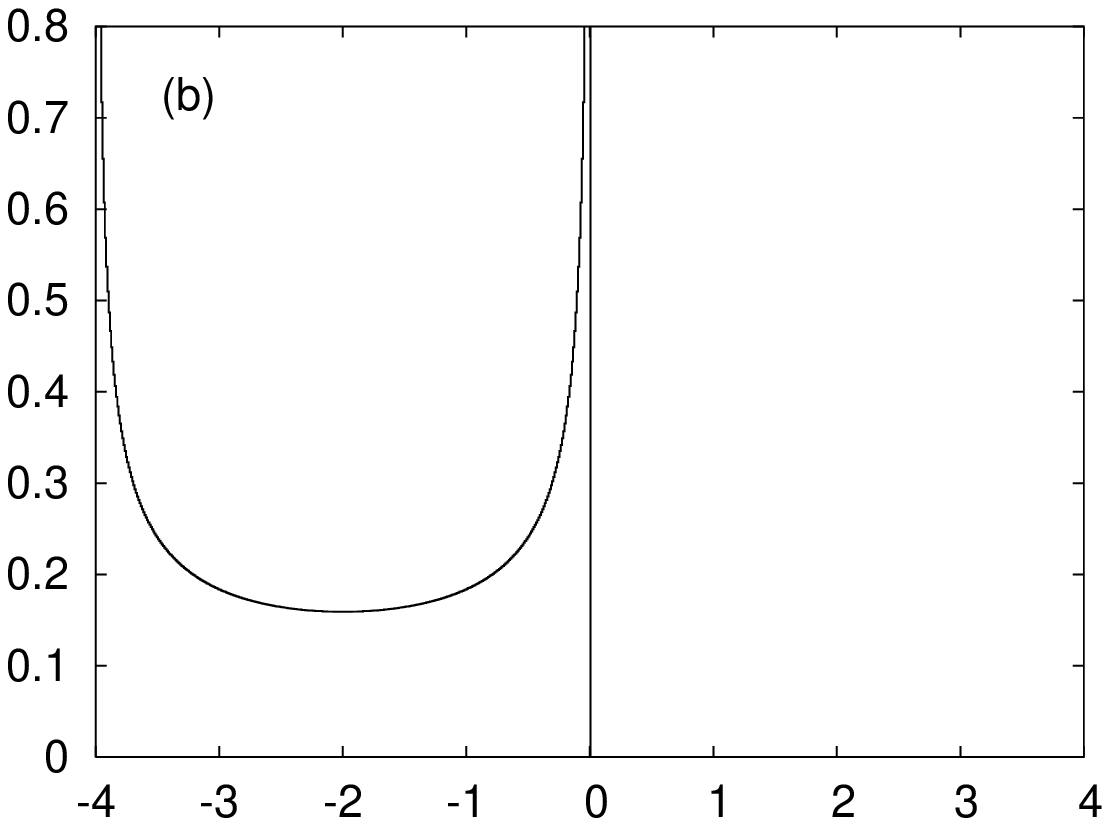}} & 
 \psfrag{(d)}[B1][B1][2][0]{(d)}\scalebox{0.33}{\includegraphics{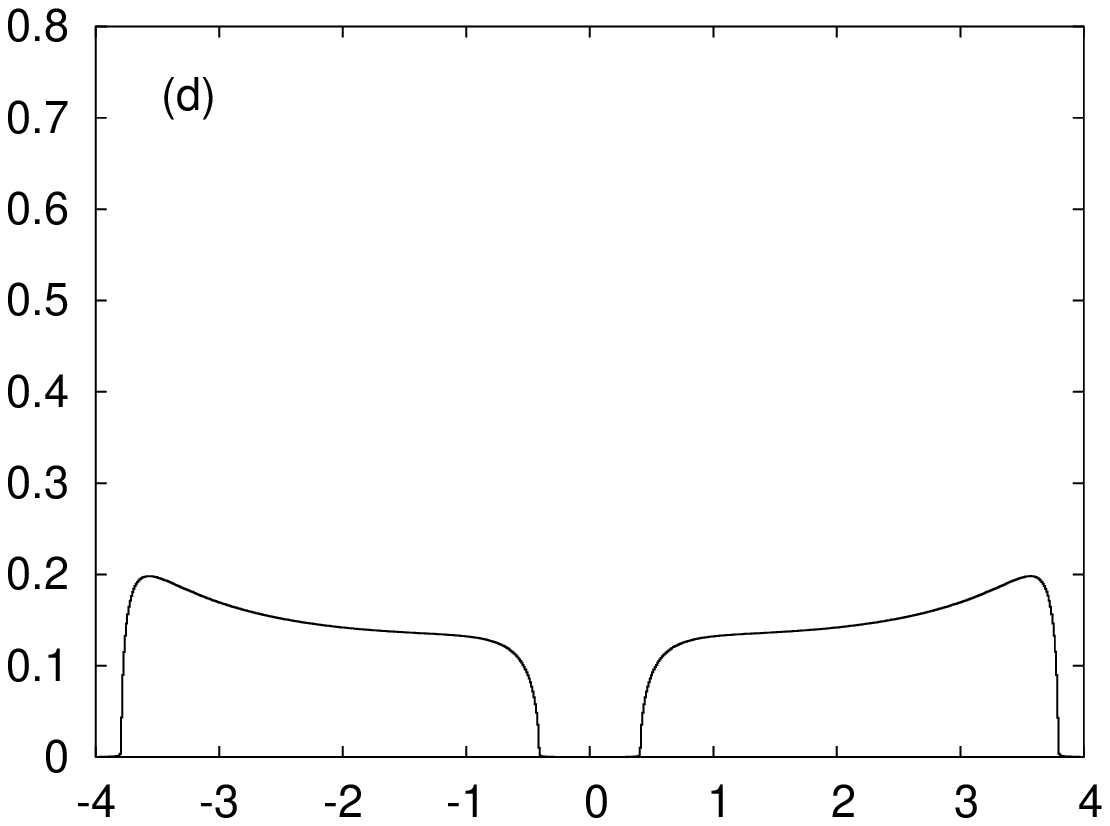}} 
 \end{tabular}         
 \caption{(a) DOS (as a function of energy) for a pure material comprising of $A$ sites, with $\epsilon_A=+2.0$. 
 					(b) DOS for a pure material comprising of $B$ sites, with $\epsilon_B=-2.0$. 
					(c) Exact DOS results for a random $A_{50}B_{50}$ alloy of the pure materials above. 
					(d) DOS for the same $A_{50}B_{50}$ alloy obtained using the CPA.}\label{pureexactcpa}
 \end{center}
\end{figure}

In this section results are presented for a 1D tight-binding model with diagonal disorder only and nearest-neighbour hopping. Calculations
for such a model were first performed with the NLCPA by Moradian \cite{Moradian1} using $N_c=4$ and $N_c=8$ clusters. Here the random site
energies can take the values $\epsilon_A=+2.0$ and $\epsilon_B=-2.0$, and the nearest neighbour hopping parameter is taken to be $W=1.0$.
Figure~\ref{pureexactcpa}(a) and figure~\ref{pureexactcpa}(b) show the DOS results for a pure material comprising of energies $\epsilon_A$ or
$\epsilon_B$ respectively. Notice that the pure bands just touch at $E=0.0$, and so a theory for an alloy of the two pure materials must deal
here with this difficult `split-band' regime. Indeed, from the Saxon-Hutner theorem \cite{Ziman1}, the gap at $E=0.0$ for an alloy of the
two pure materials should correspondingly be just vanishing. 

\subsection{Exact Result and the CPA}

Figure~\ref{pureexactcpa}(c) shows exact DOS results for a random $A_{50}B_{50}$ alloy of the two pure materials shown in
figures~\ref{pureexactcpa}(a),(b) obtained using the negative eigenvalue theorem \cite{Dean1}. The results have been displayed using centred
histograms at 400 points from $E=-4.0$ to $E=+4.0$ i.e.~an energy increment of 0.01 (a higher resolution yields too much structure for
meaningful comparisons to be made). As expected, the DOS is highly structured and the gap at $E=0.0$ is just vanishing. 

Next, figure~\ref{pureexactcpa}(d) shows a CPA calculation for the same $A_{50}B_{50}$ alloy, displayed using centred histograms at the same
energy resolution as the exact result. The CPA result is very smooth due to the neglect of nonlocal fluctuations in the disorder
configurations. However, one of the central successes of the CPA is its ability to describe the split-band regime. Indeed the CPA correctly
separates the DOS here into two sub-bands, but it is clear that it fails to correctly describe the onset of such split-band behaviour. This
is because the contributions to the DOS near band edges are due to large clusters of like atoms which are not described by the CPA
\cite{Gonis1}. 

\subsection{Cluster theories}

\begin{figure*}[!]
 \begin{center}
 \begin{tabular}{ccc}
   \psfrag{ECM}[B1][B1][2][0]{ECM}\scalebox{0.33}{\includegraphics{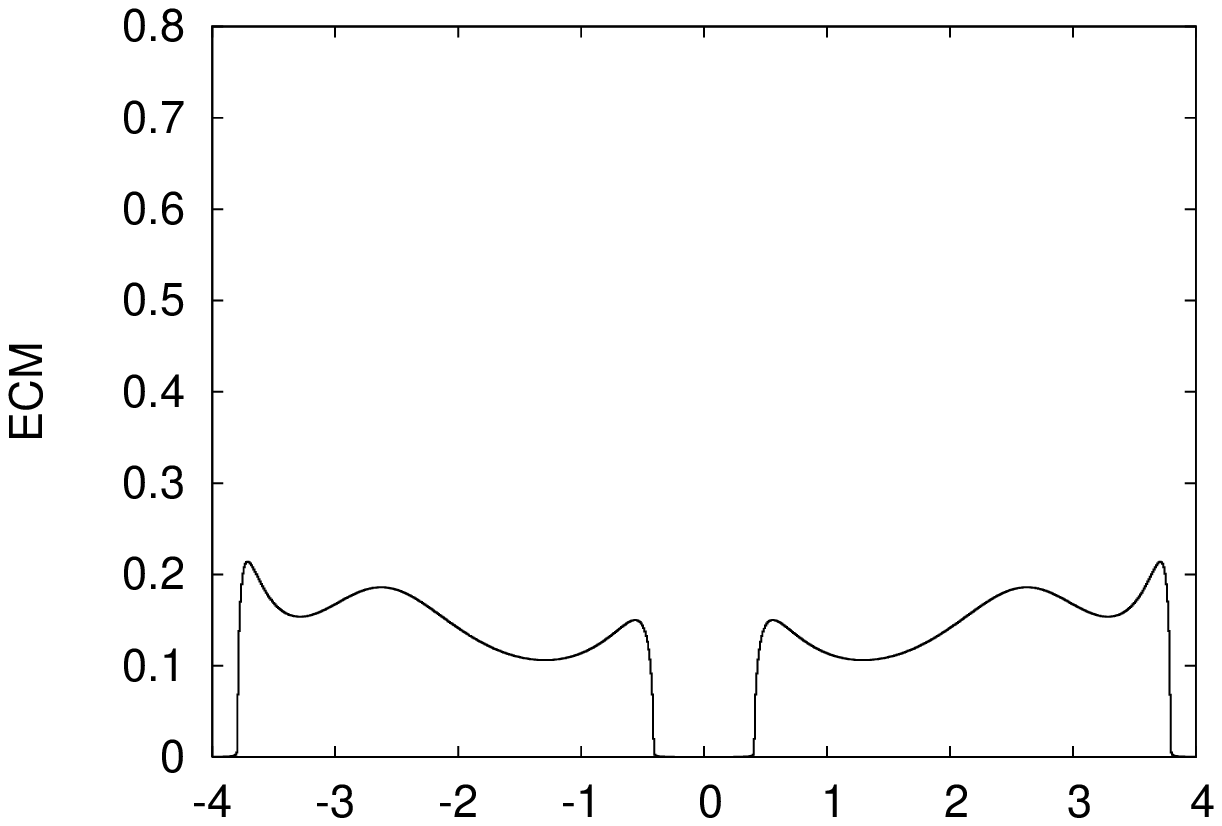}}     
 & \scalebox{0.33}{\includegraphics{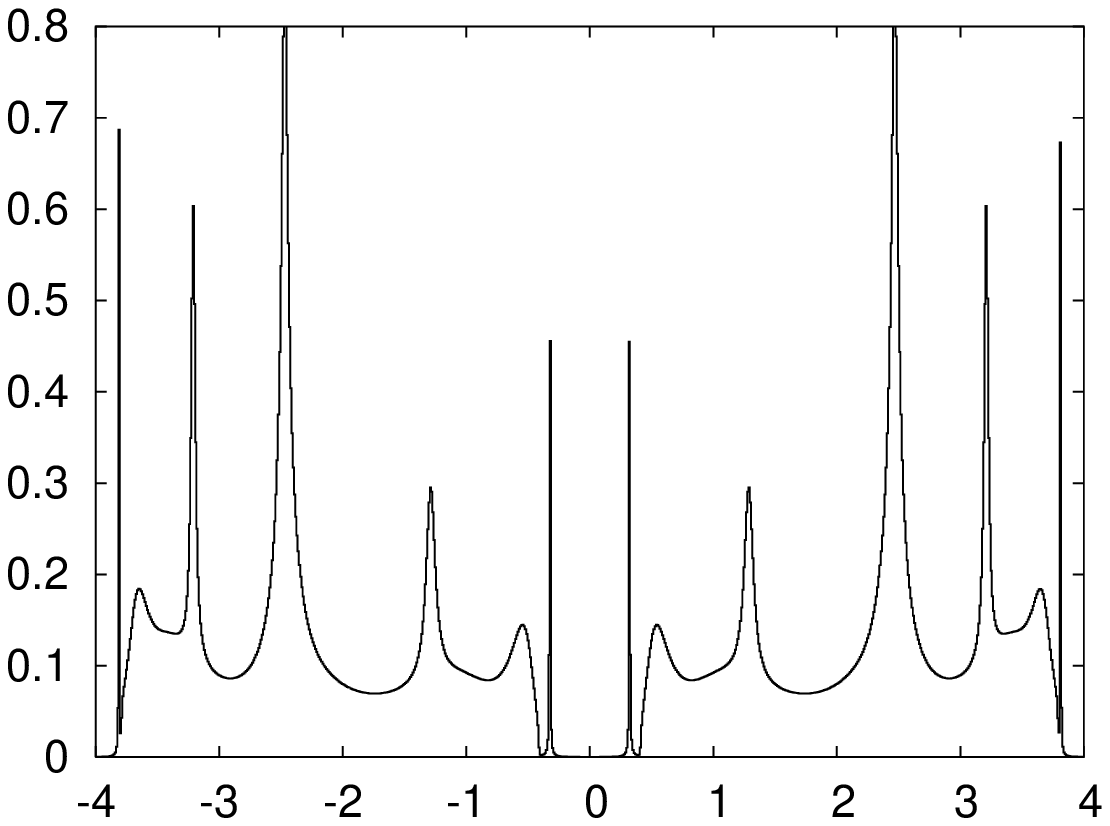}}     
 & \scalebox{0.33}{\includegraphics{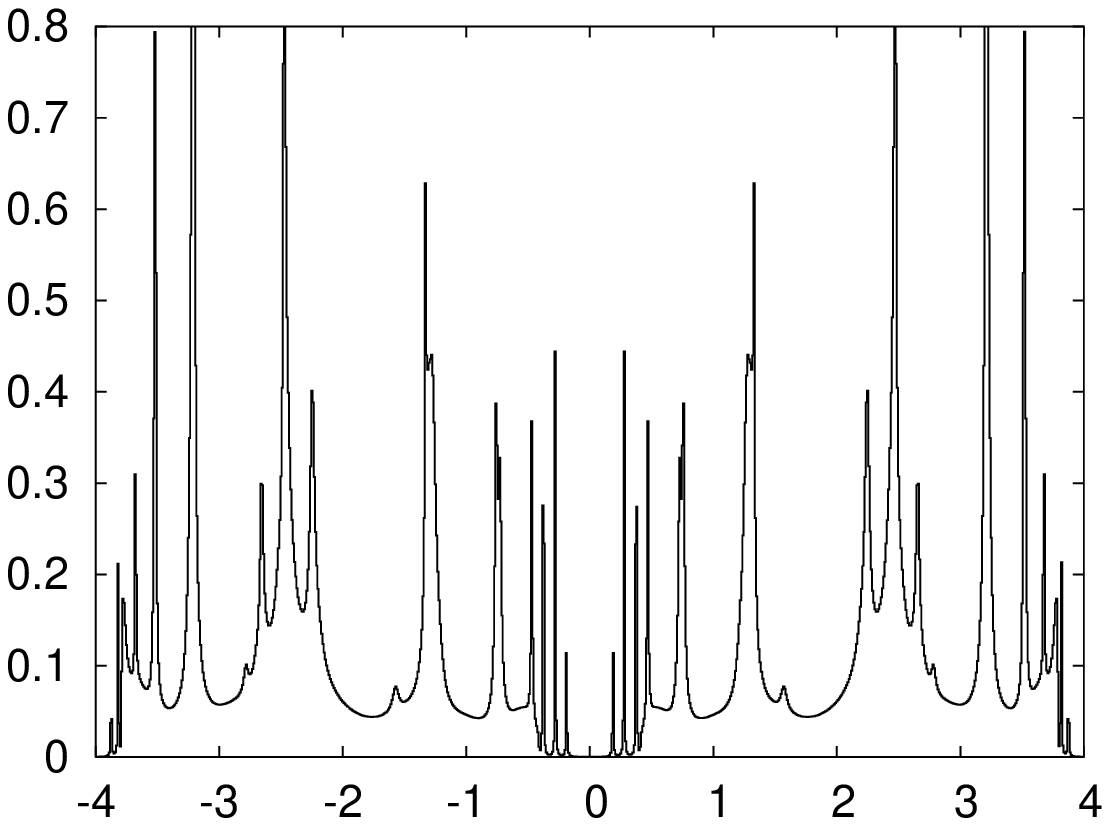}}\\
   \psfrag{MCPA}[B1][B1][2][0]{MCPA centre}\scalebox{0.33}{\includegraphics{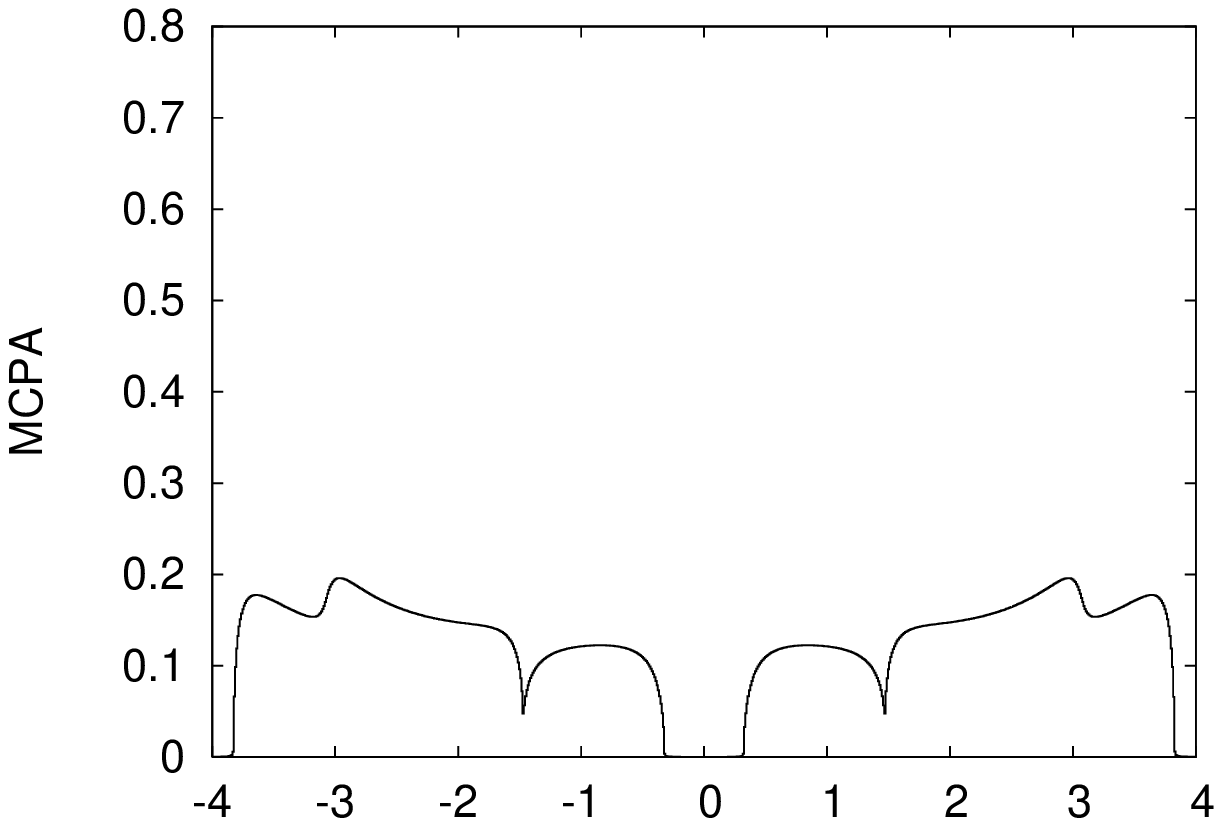}}    
 & \scalebox{0.33}{\includegraphics{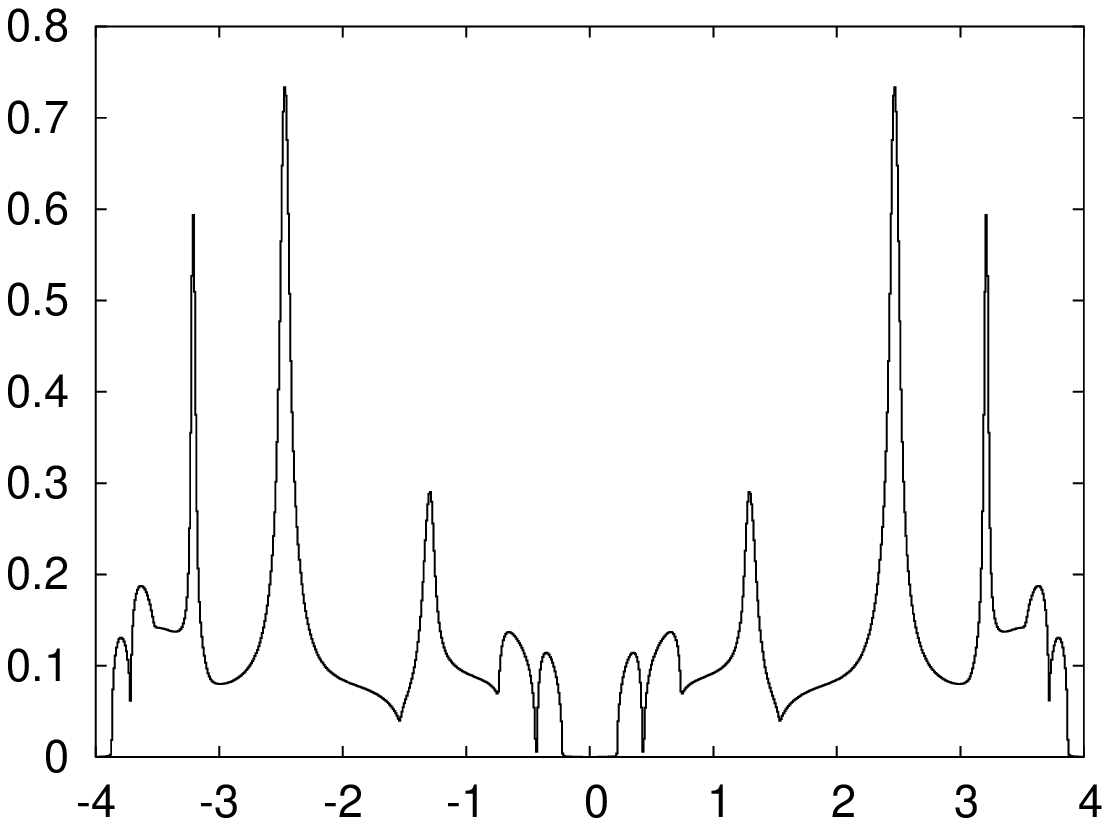}}   
 & \scalebox{0.33}{\includegraphics{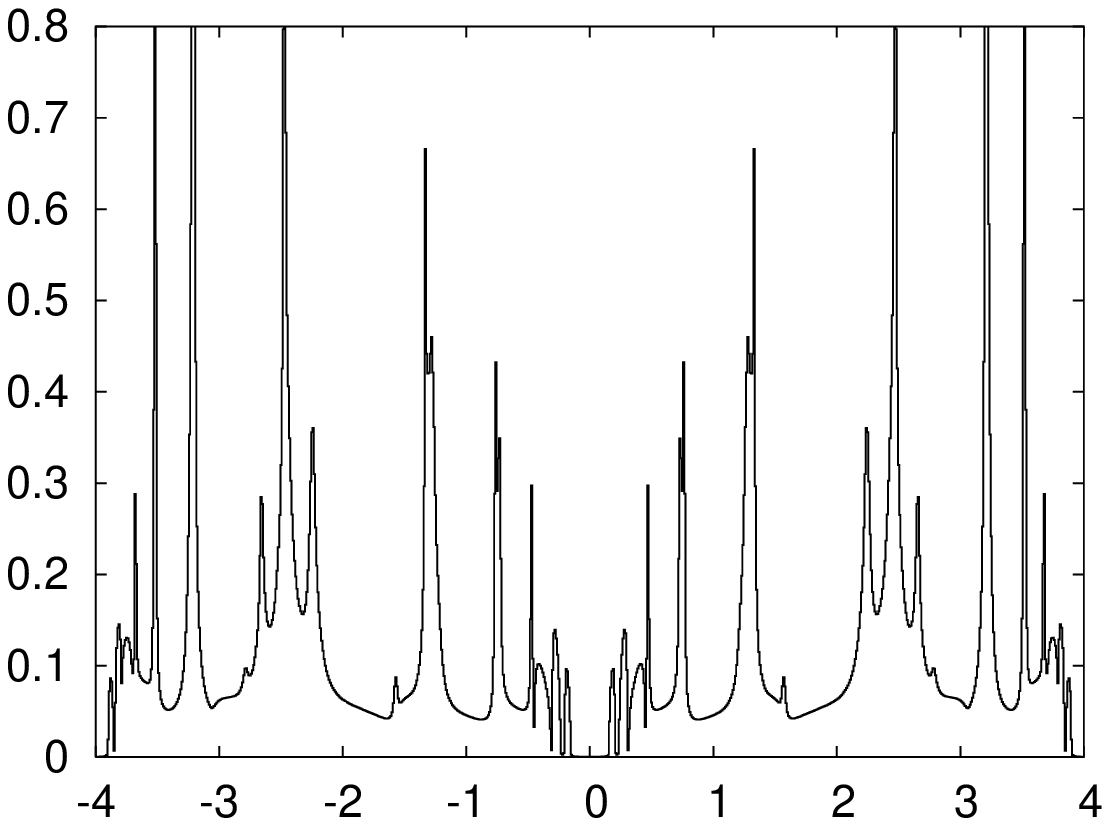}}\\ 
   \psfrag{MCPA}[B1][B1][2][0]{MCPA average}\scalebox{0.33}{\includegraphics{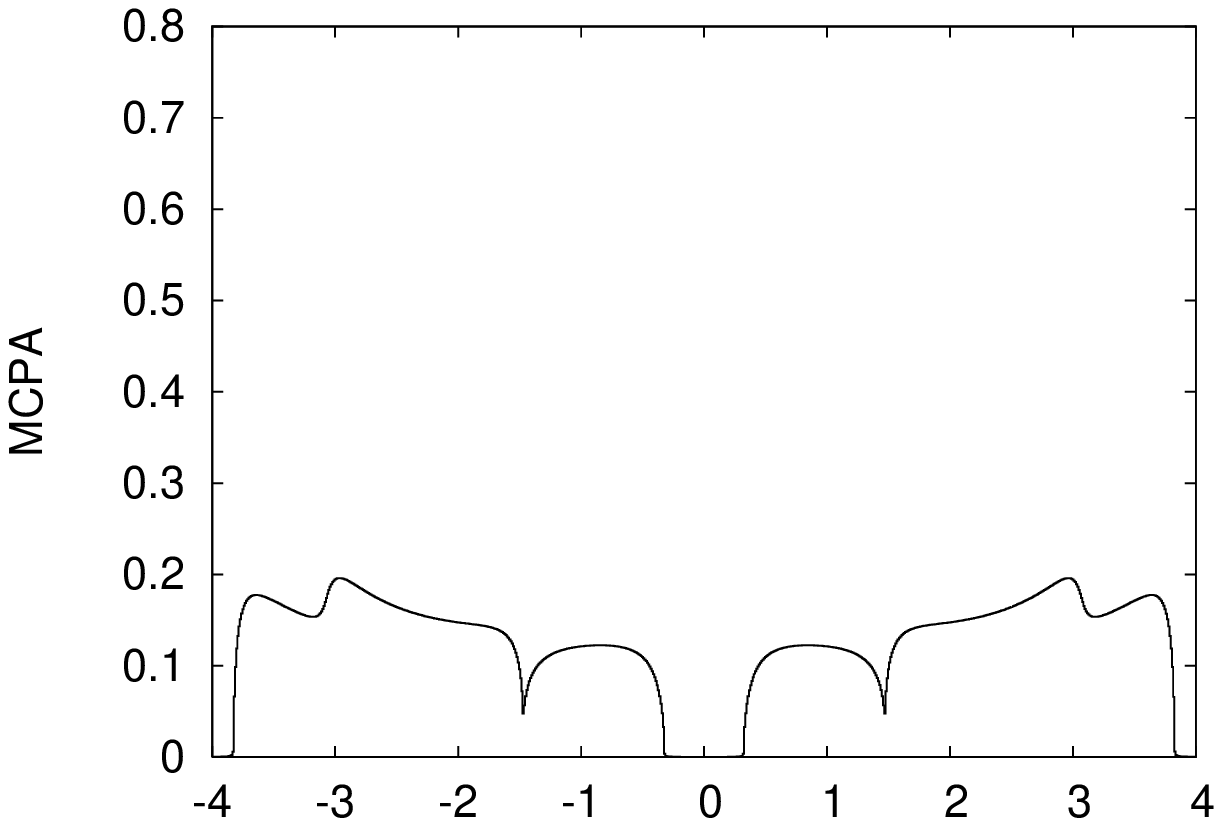}} 
 & \scalebox{0.33}{\includegraphics{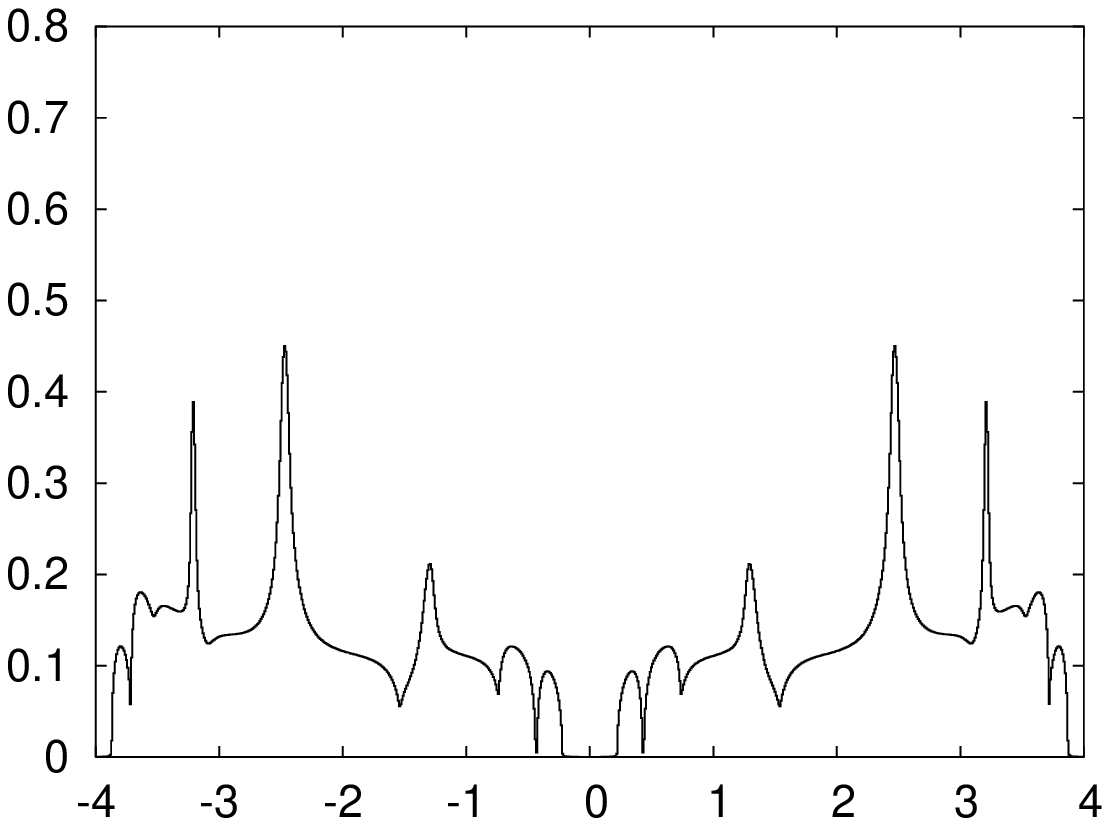}} 
 & \scalebox{0.33}{\includegraphics{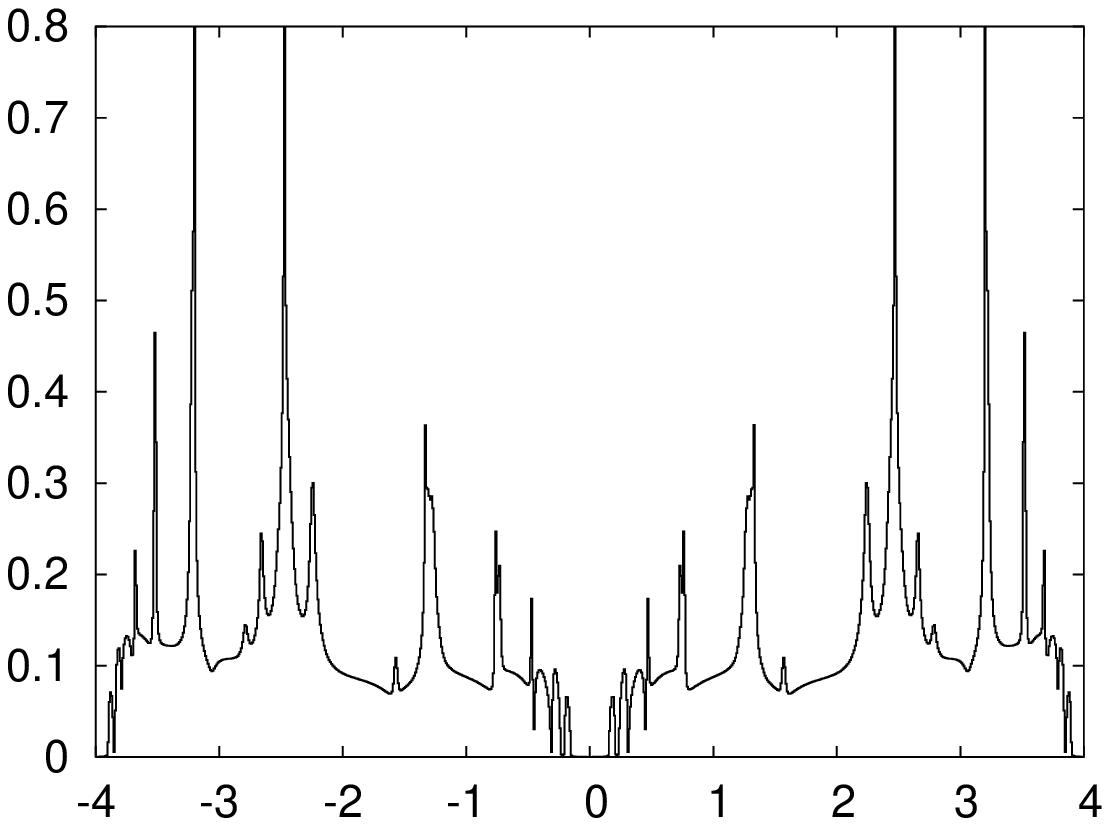}}\\
   \psfrag{NLCPA}[B1][B1][2][0]{NLCPA periodic}\scalebox{0.33}{\includegraphics{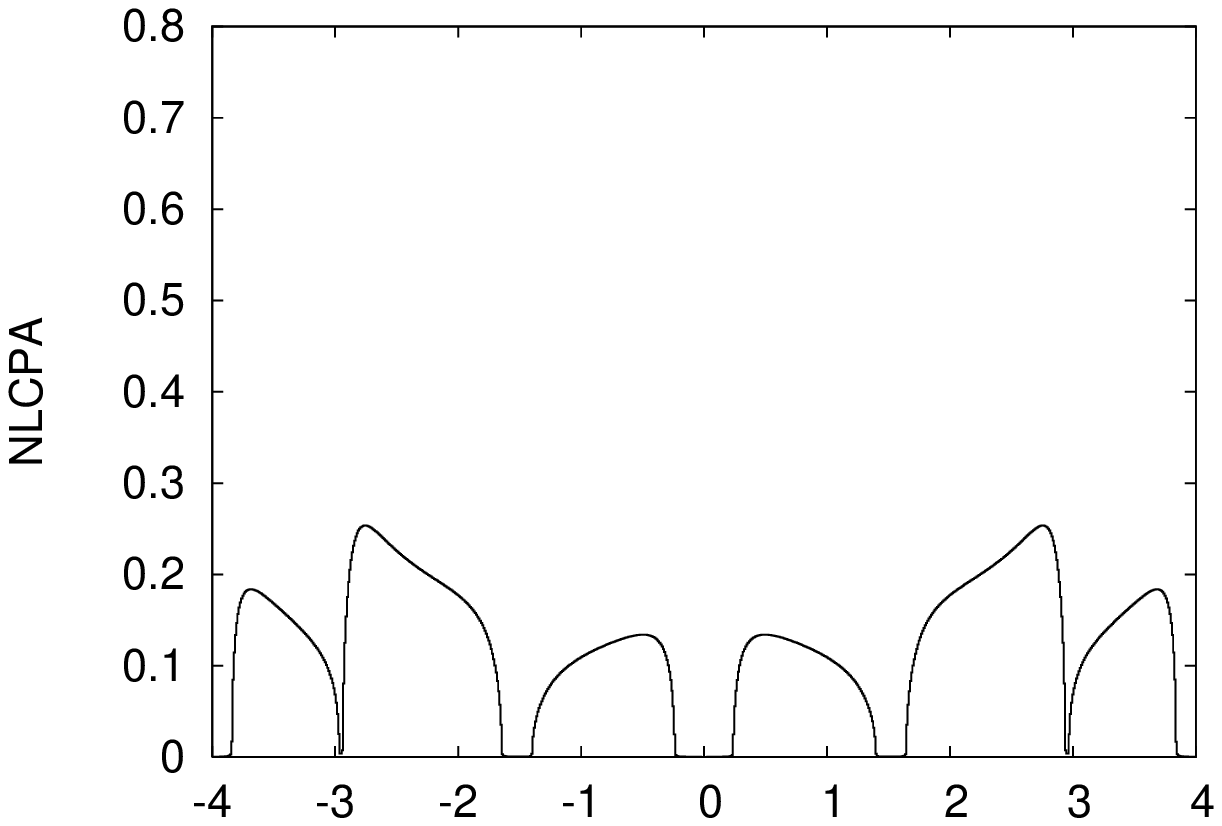}}
 & \scalebox{0.33}{\includegraphics{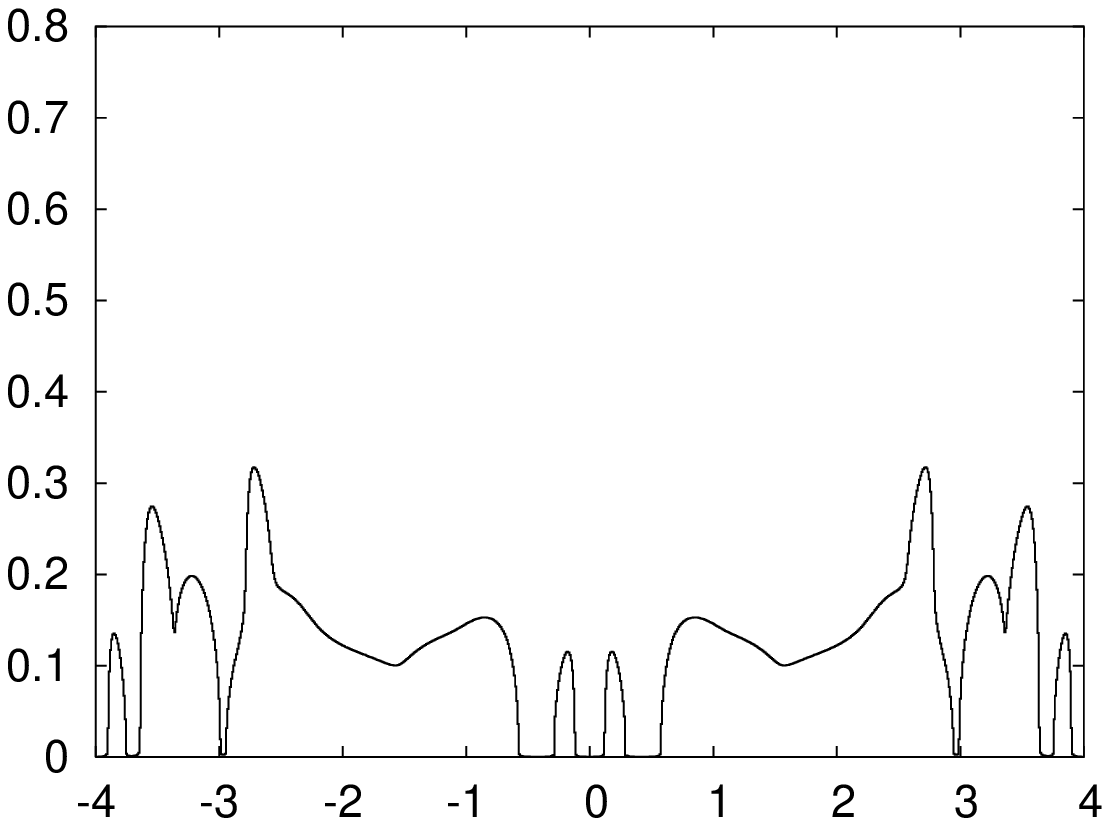}}     
 & \scalebox{0.33}{\includegraphics{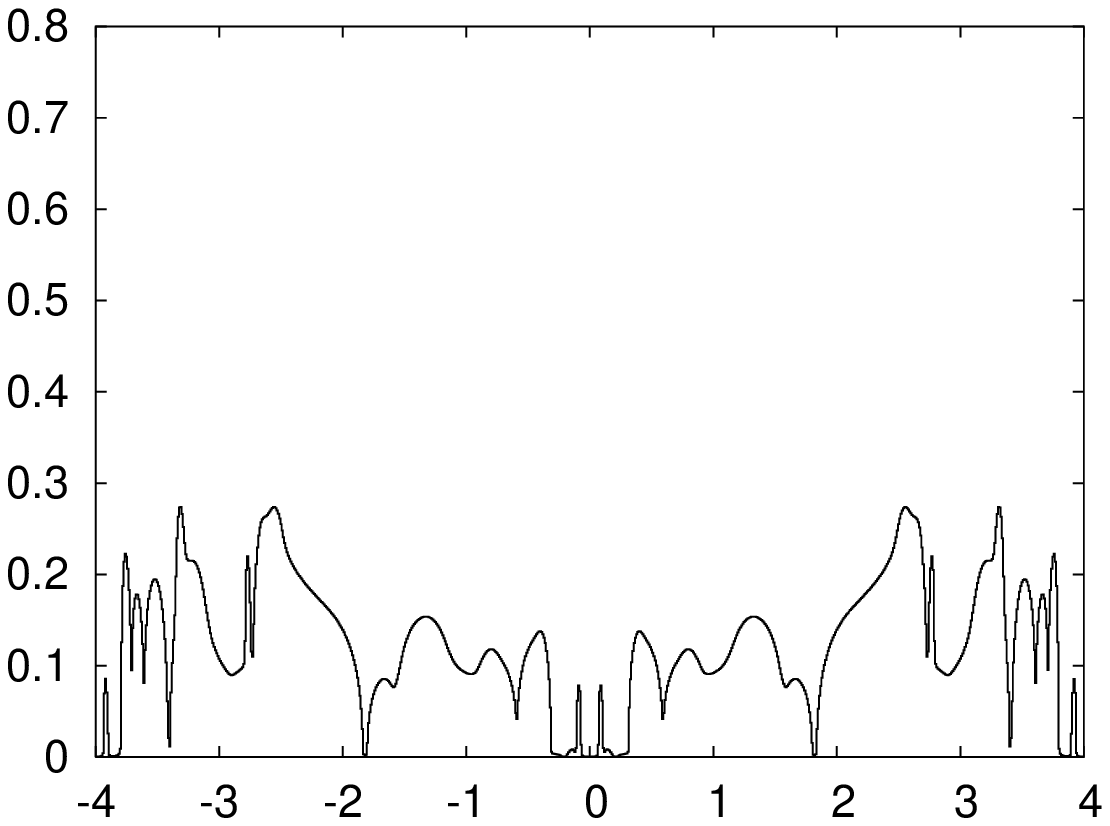}}\\
   \psfrag{NLCPA}[B1][B1][2][0]{NLCPA anti-periodic}\scalebox{0.33}{\includegraphics{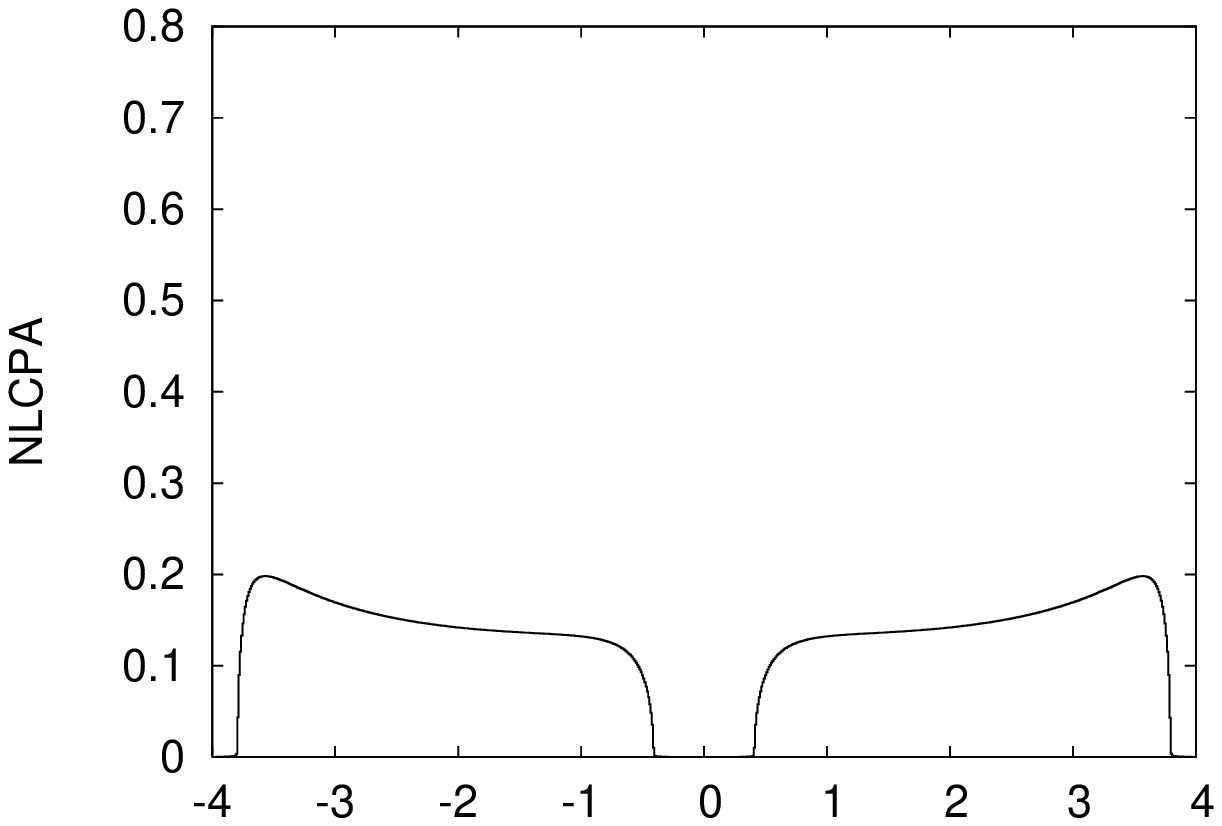}} 
 & \scalebox{0.33}{\includegraphics{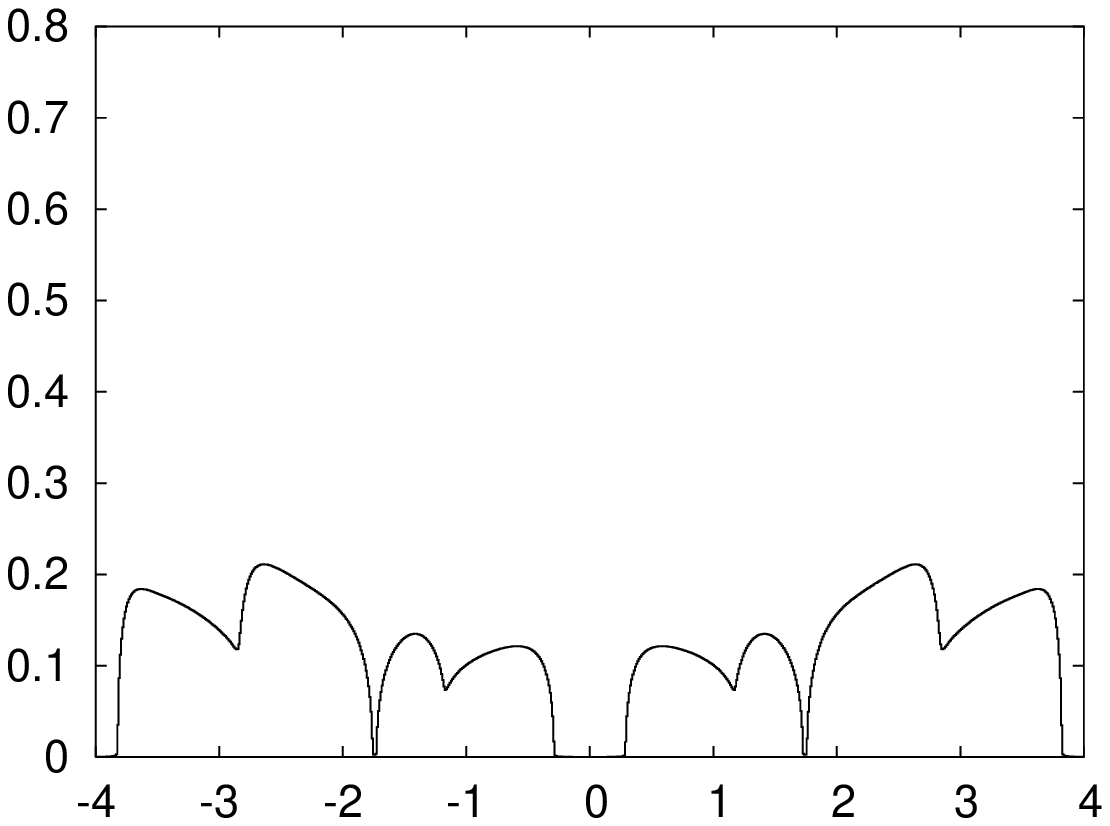}}     
 & \scalebox{0.33}{\includegraphics{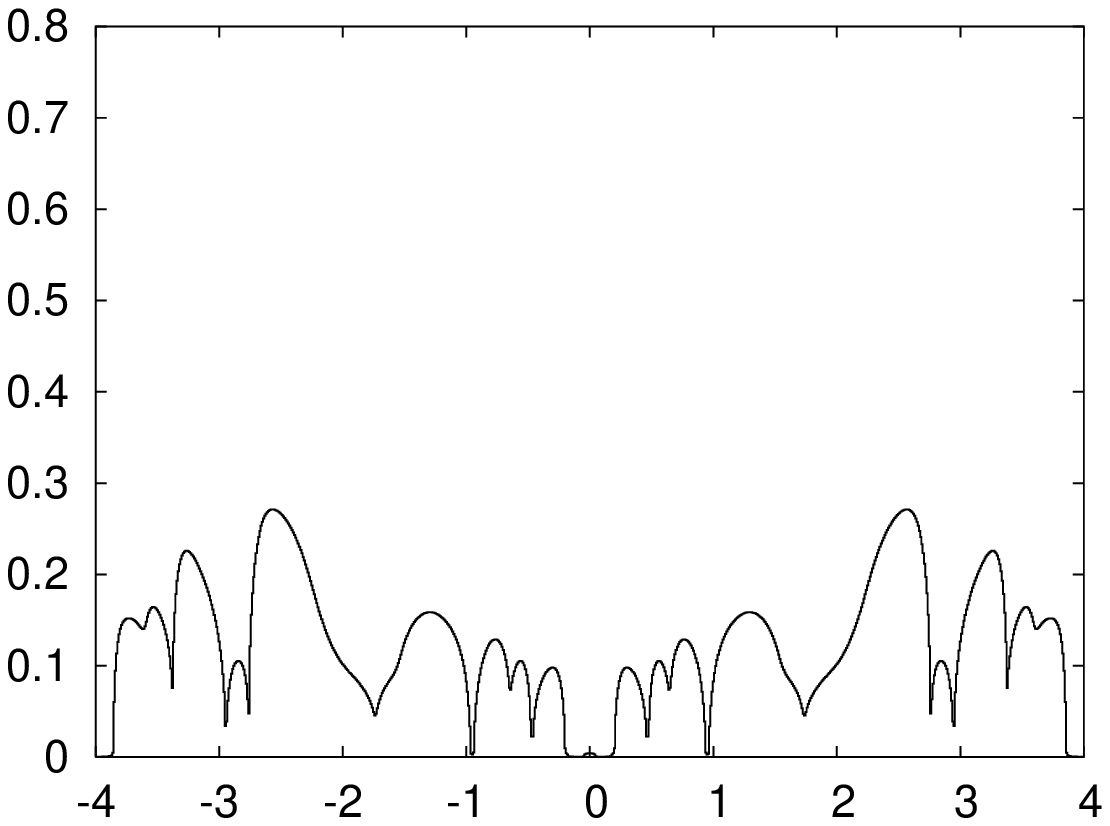}}\\ 
   \psfrag{Nc2}[B1][B1][2.5][0]{$N_C=2$}\psfrag{NLCPA}[B1][B1][2][0]{NLCPA average}\scalebox{0.33}{\includegraphics{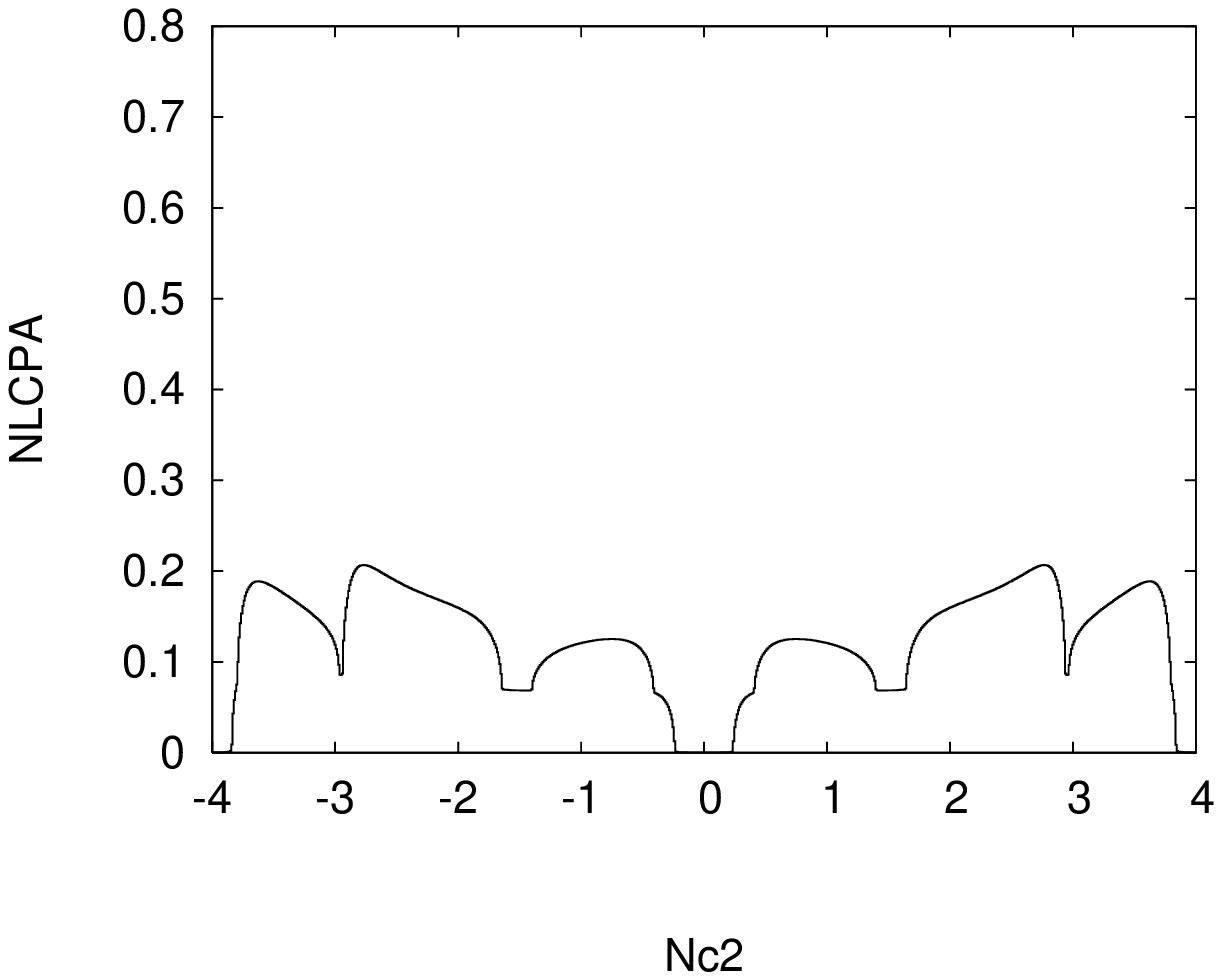}}
 & \psfrag{Nc4}[B1][B1][2.5][0]{$N_C=4$}\scalebox{0.33}{\includegraphics{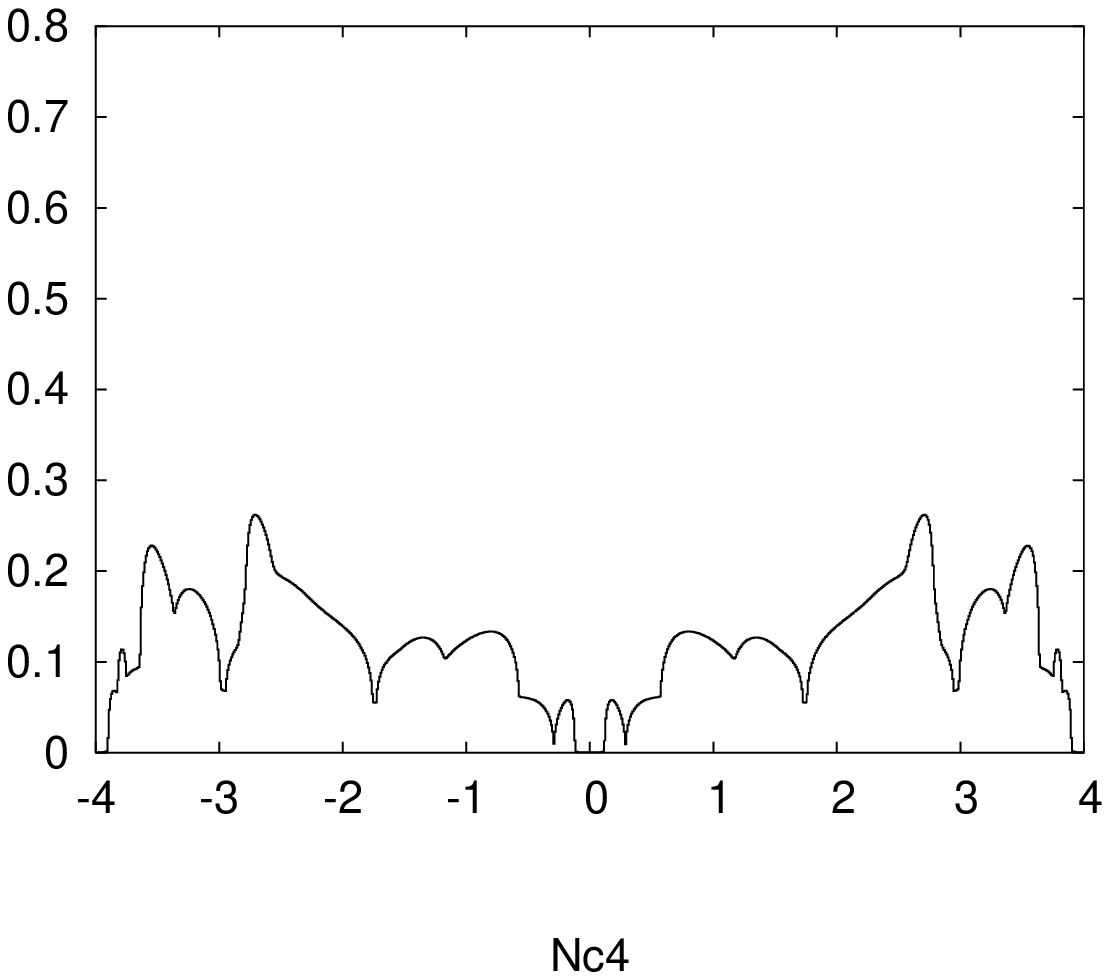}}   
 & \psfrag{Nc6}[B1][B1][2.5][0]{$N_C=6$}\scalebox{0.33}{\includegraphics{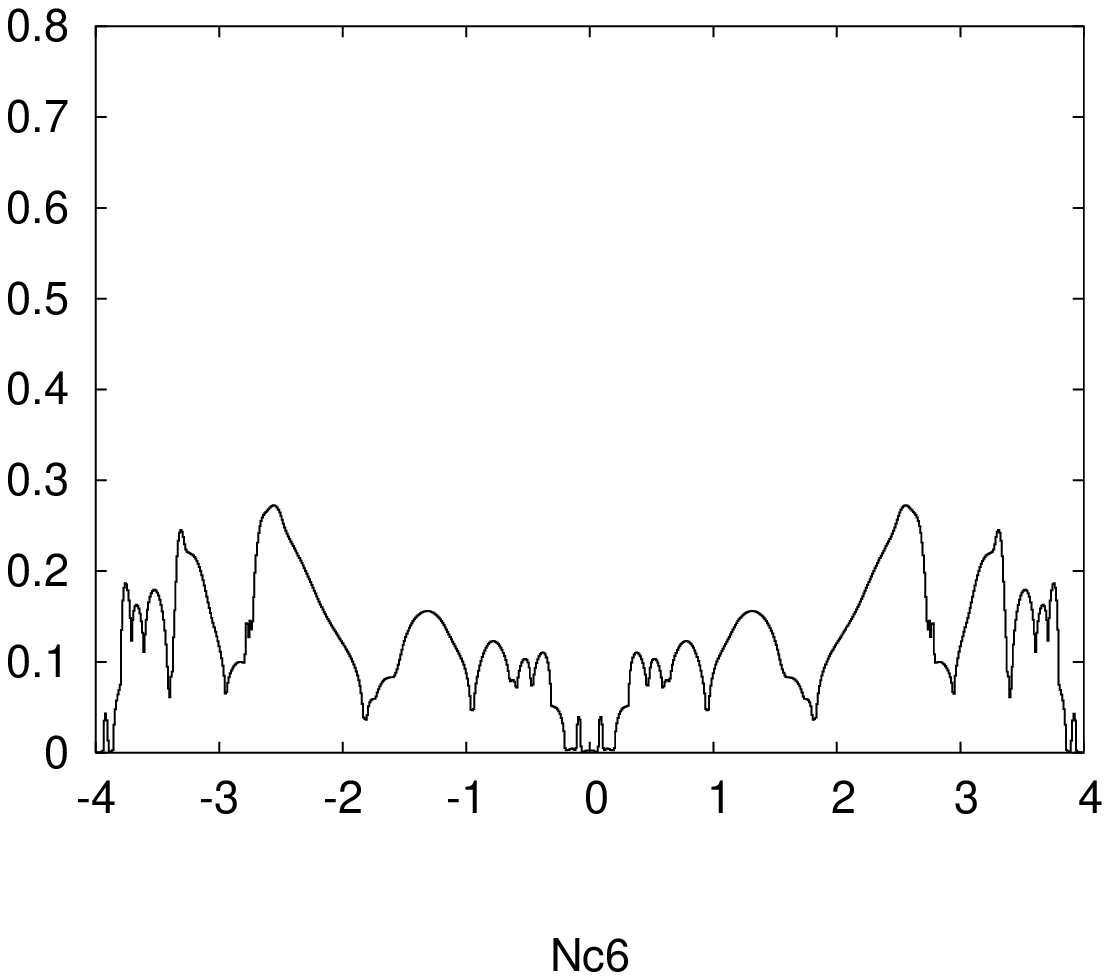}}    
 \end{tabular}         
 \caption{Configurationally-averaged DOS per site as a function of energy (in units of the bandwidth) for the various cluster theories (top 
 to bottom) with cluster sizes $N_c=2,4,6$ (left to right).}\label{cluster246}
 \end{center}
\end{figure*}

\begin{figure*}[!]
 \begin{center}
 \begin{tabular}{ccc}
   \psfrag{ECM}[B1][B1][2][0]{ECM}\scalebox{0.33}{\includegraphics{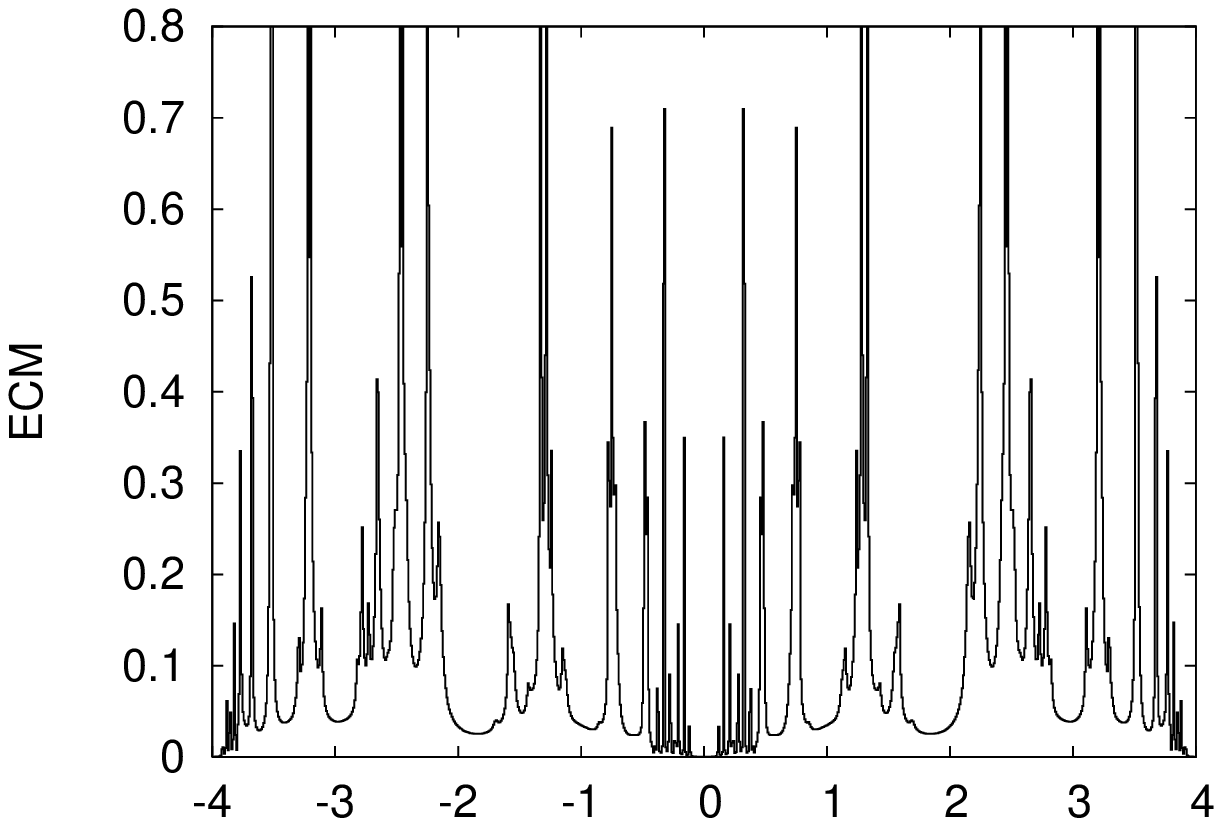}}     
 & \scalebox{0.33}{\includegraphics{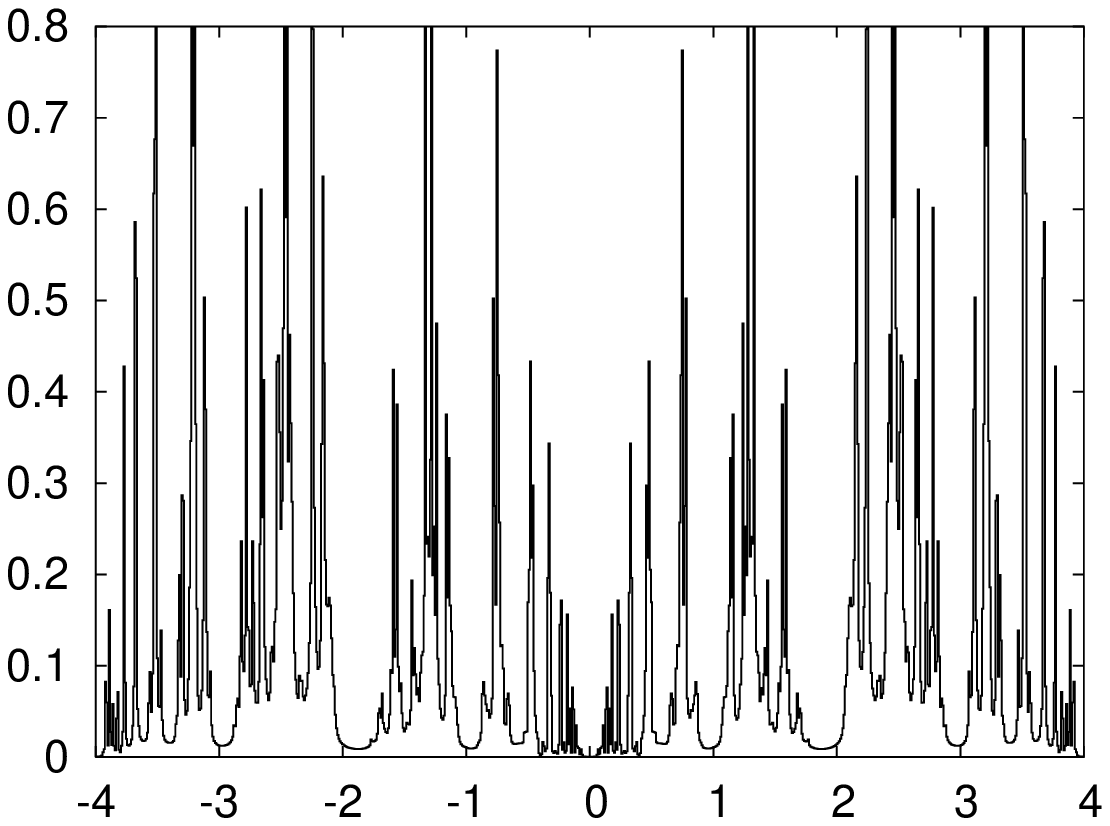}}     
 & \scalebox{0.33}{\includegraphics{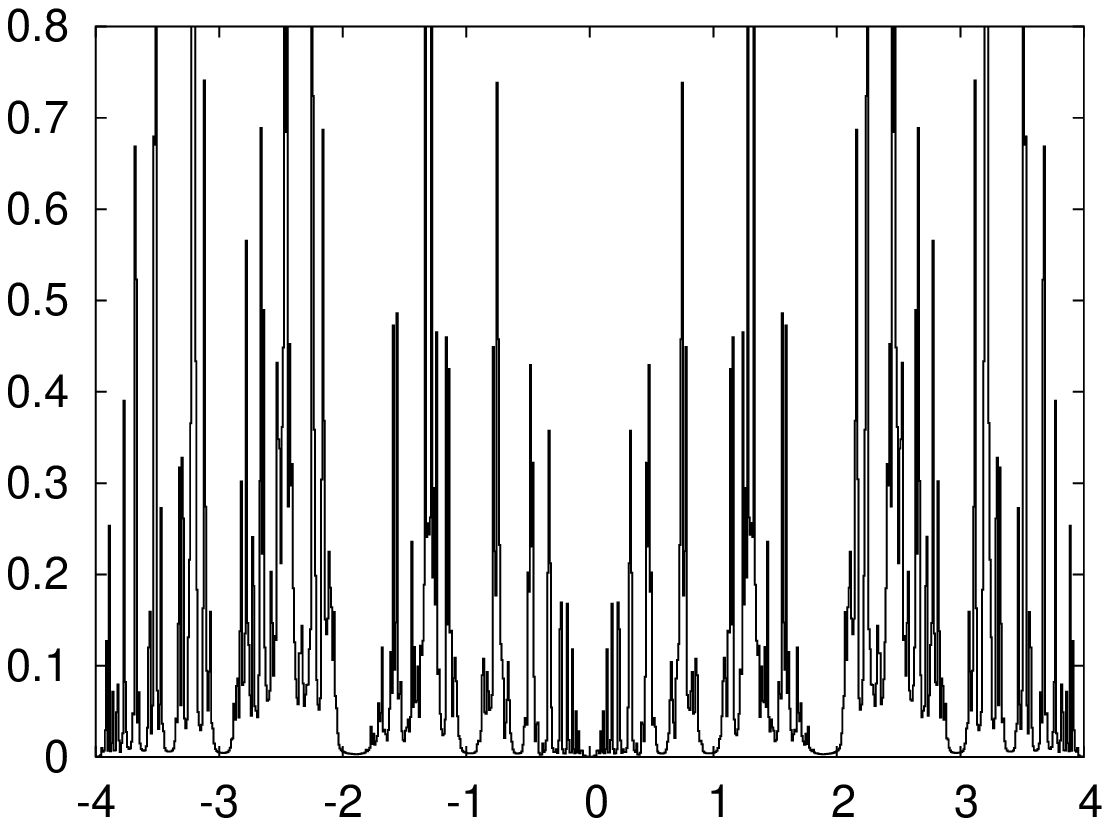}}\\
   \psfrag{MCPA}[B1][B1][2][0]{MCPA centre}\scalebox{0.33}{\includegraphics{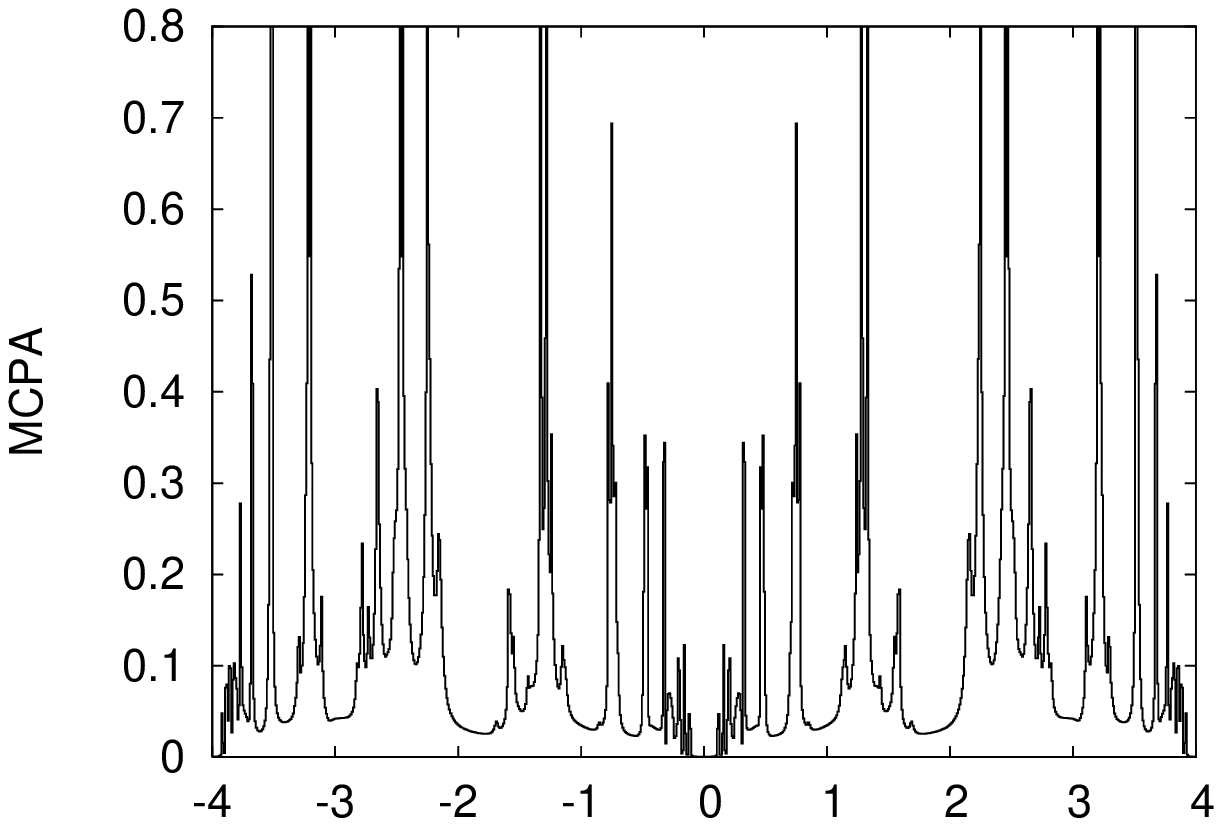}}    
 & \scalebox{0.33}{\includegraphics{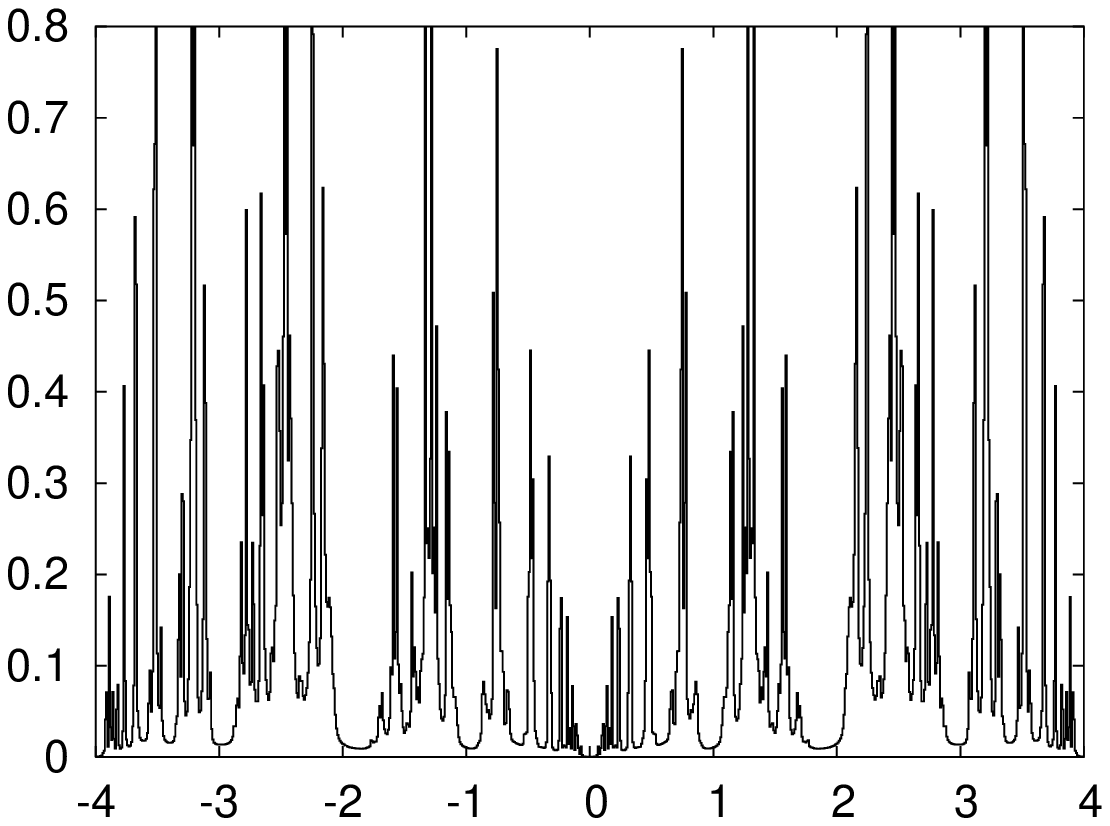}}   
 & \scalebox{0.33}{\includegraphics{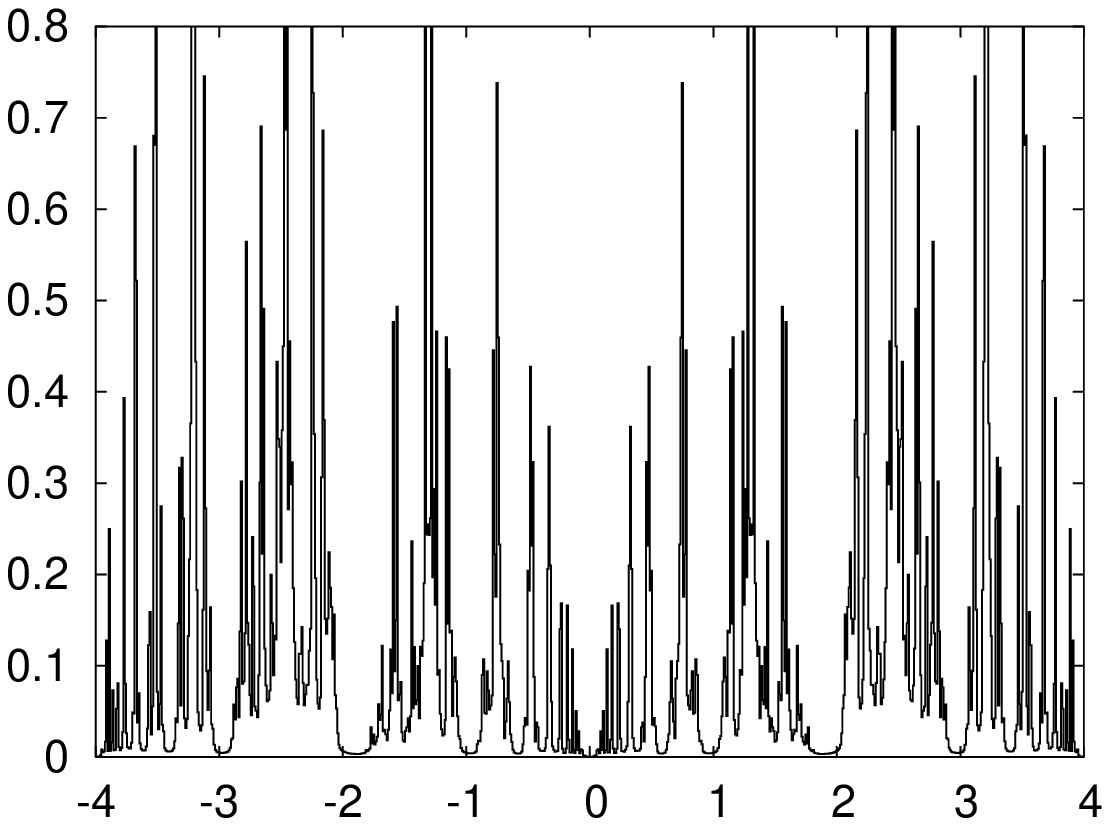}}\\ 
   \psfrag{MCPA}[B1][B1][2][0]{MCPA average}\scalebox{0.33}{\includegraphics{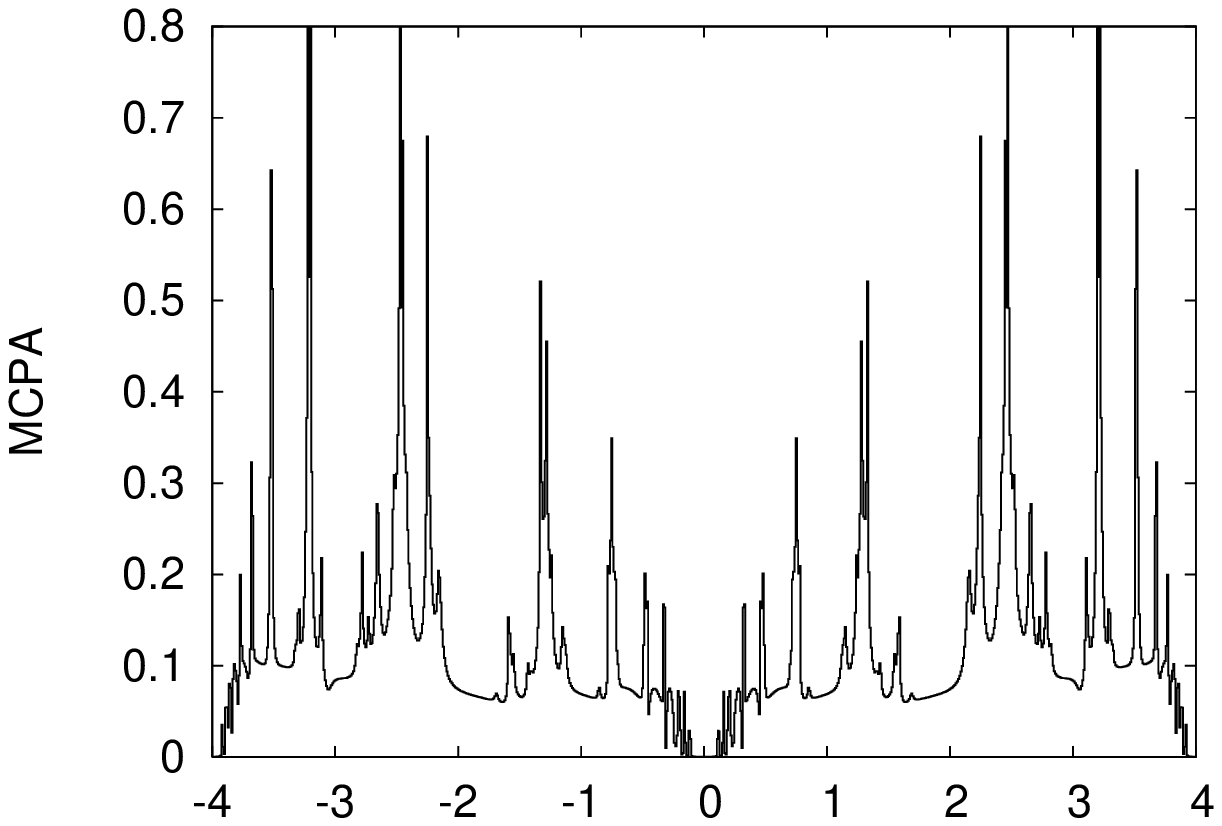}} 
 & \scalebox{0.33}{\includegraphics{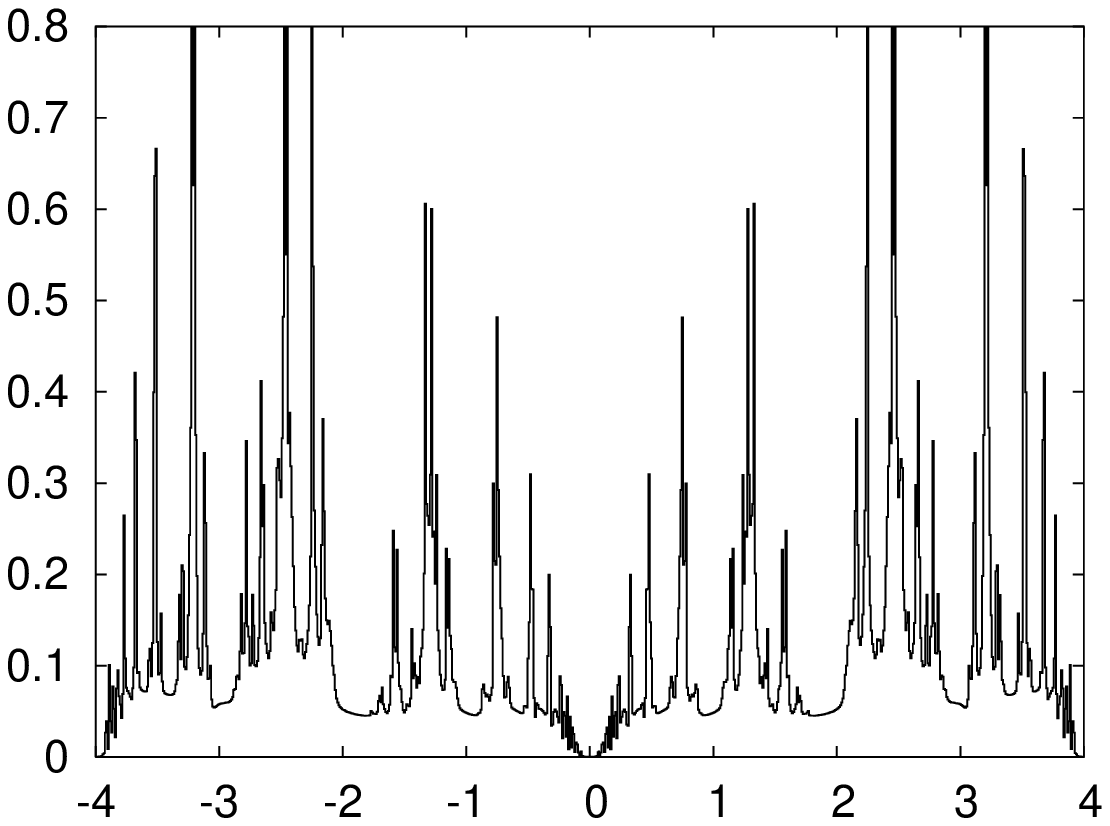}} 
 & \scalebox{0.33}{\includegraphics{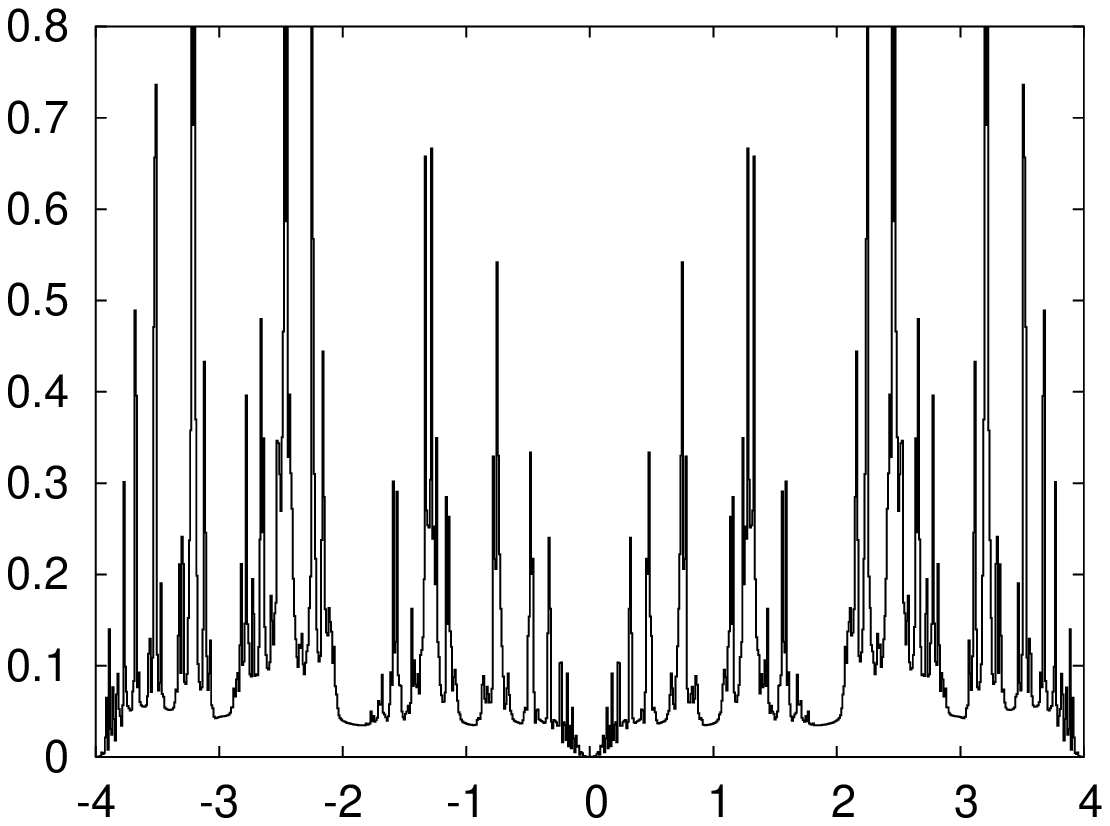}}\\
   \psfrag{NLCPA}[B1][B1][2][0]{NLCPA periodic}\scalebox{0.33}{\includegraphics{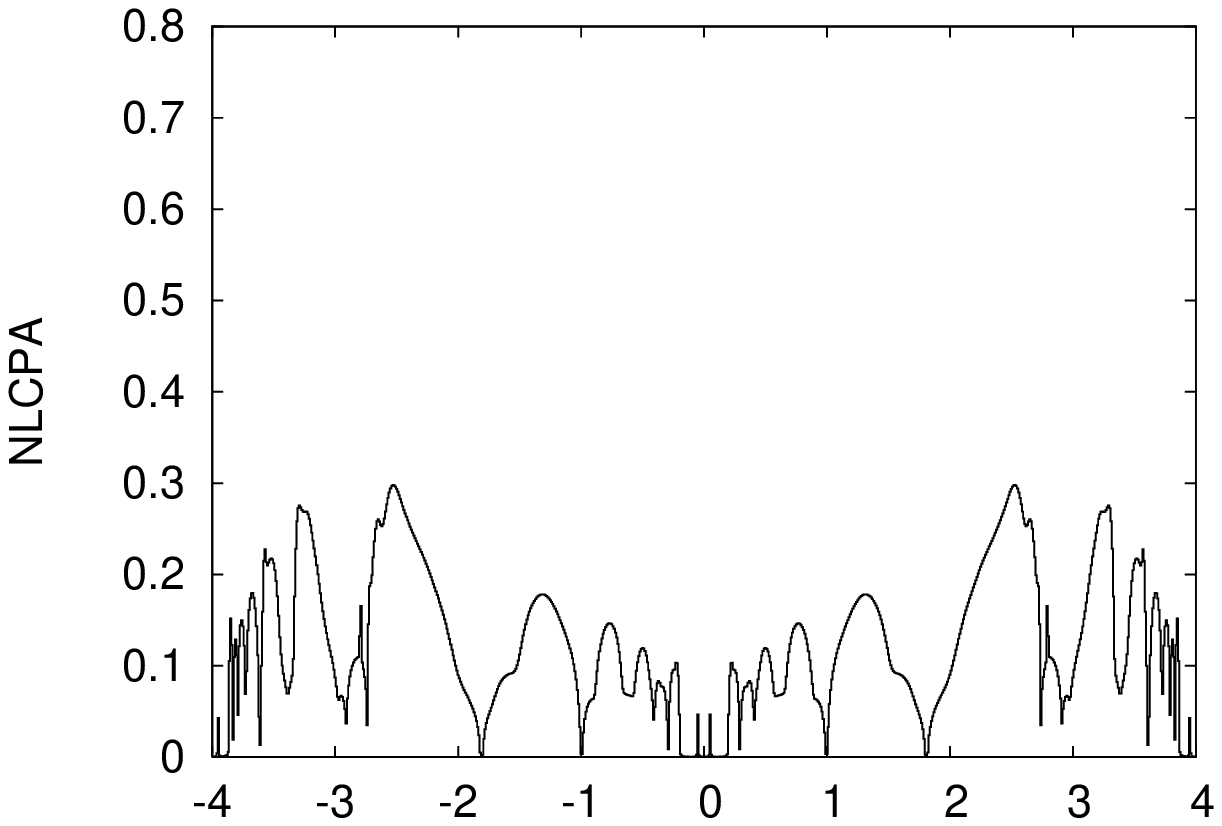}}
 & \scalebox{0.33}{\includegraphics{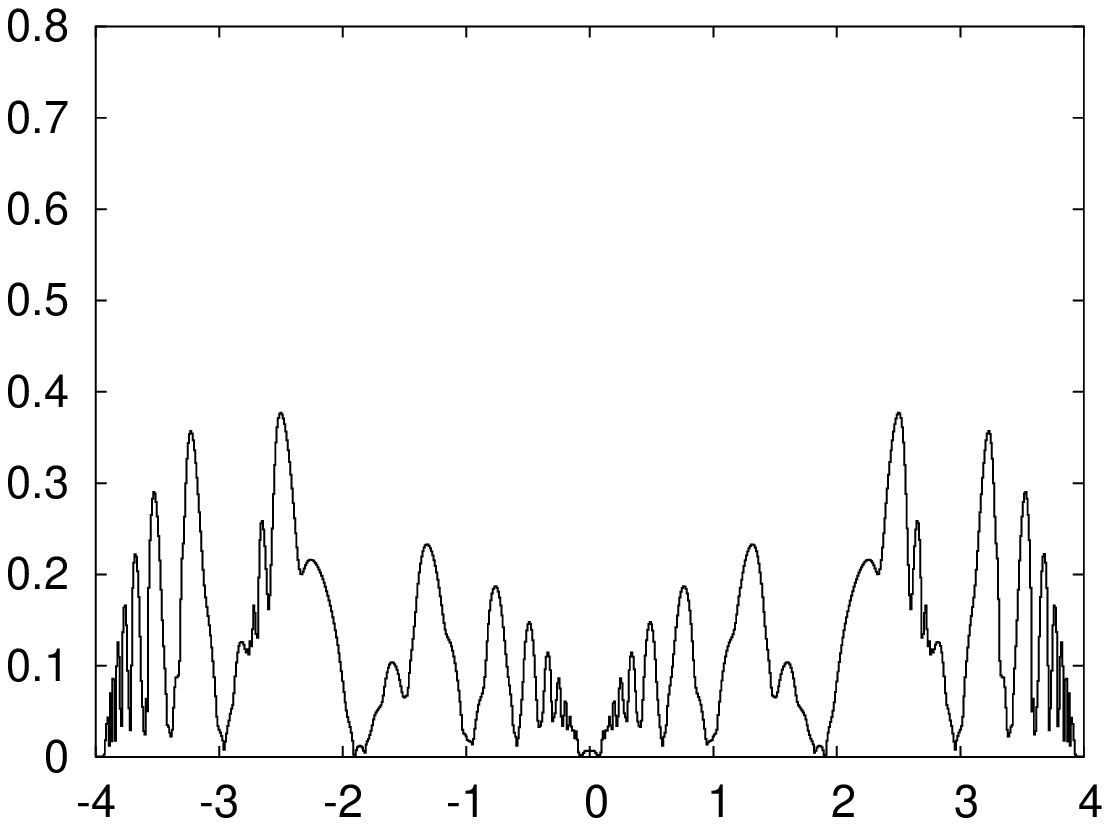}}     
 & \scalebox{0.33}{\includegraphics{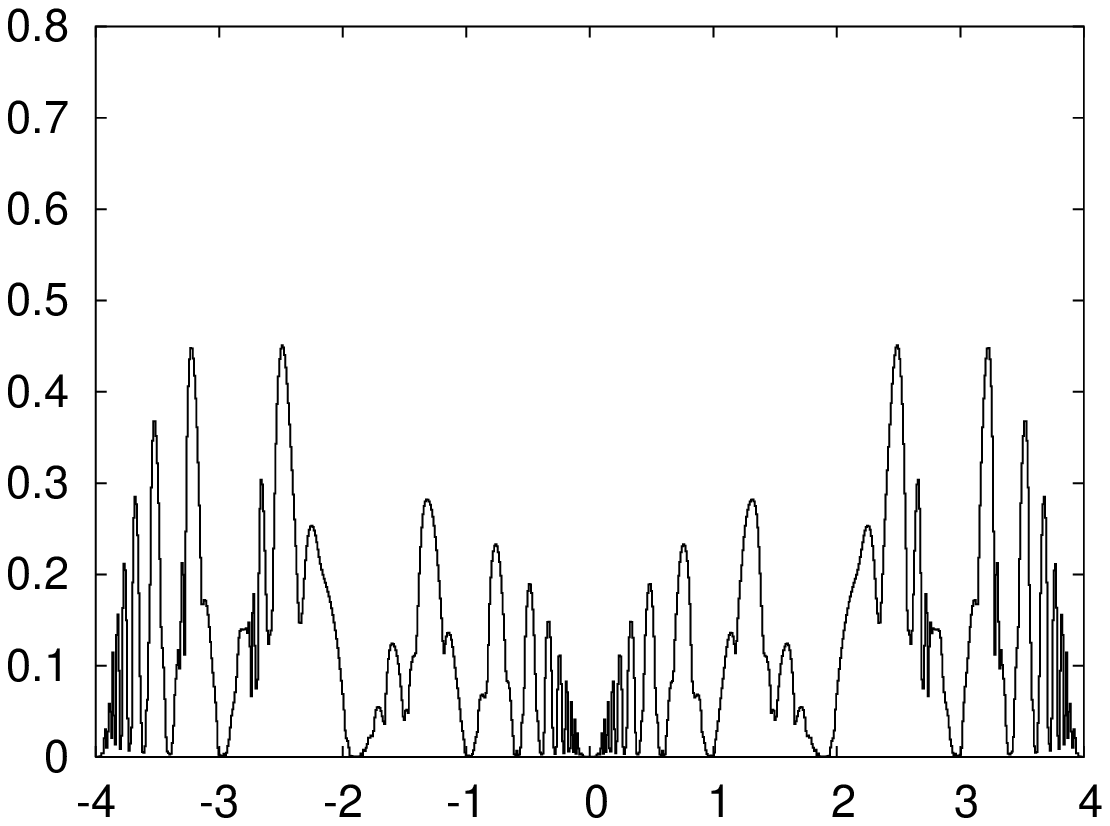}}\\
   \psfrag{NLCPA}[B1][B1][2][0]{NLCPA anti-periodic}\scalebox{0.33}{\includegraphics{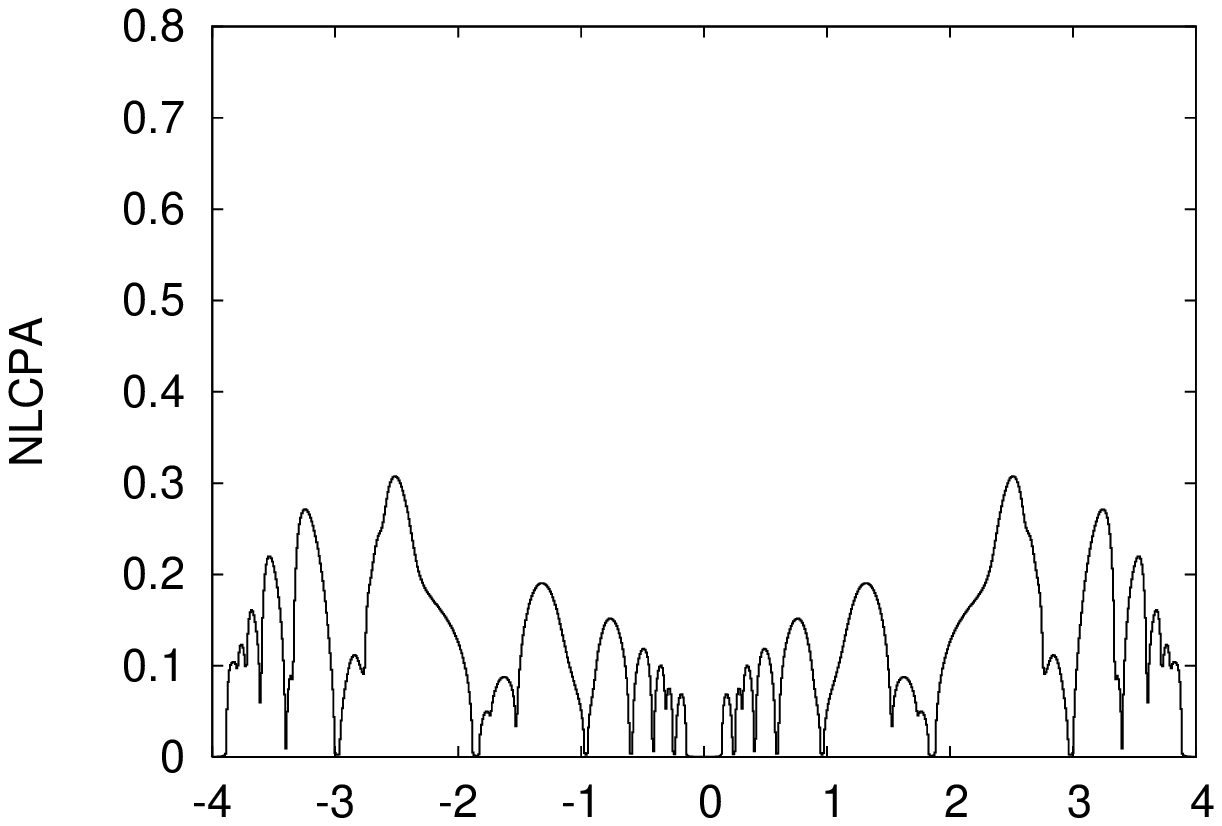}} 
 & \scalebox{0.33}{\includegraphics{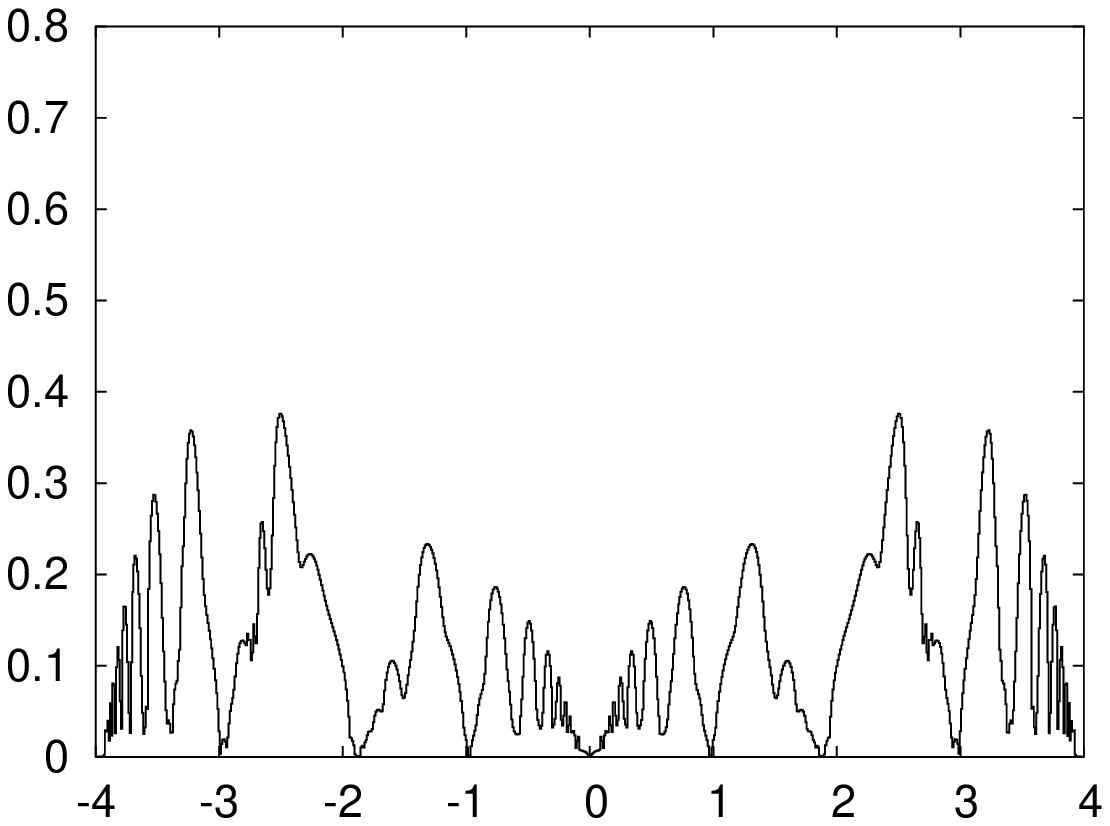}}     
 & \scalebox{0.33}{\includegraphics{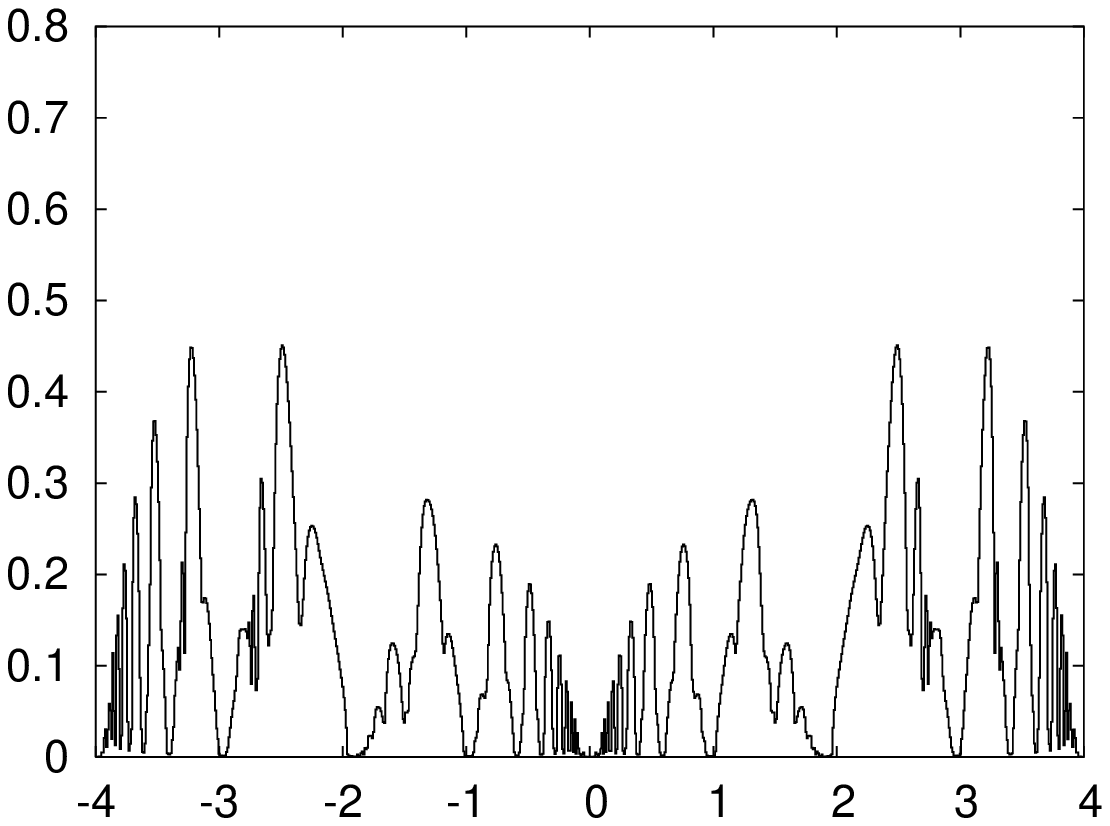}}\\ 
   \psfrag{Nc8}[B1][B1][2.5][0]{$N_C=8$}\psfrag{NLCPA}[B1][B1][2][0]{NLCPA average}\scalebox{0.33}{\includegraphics{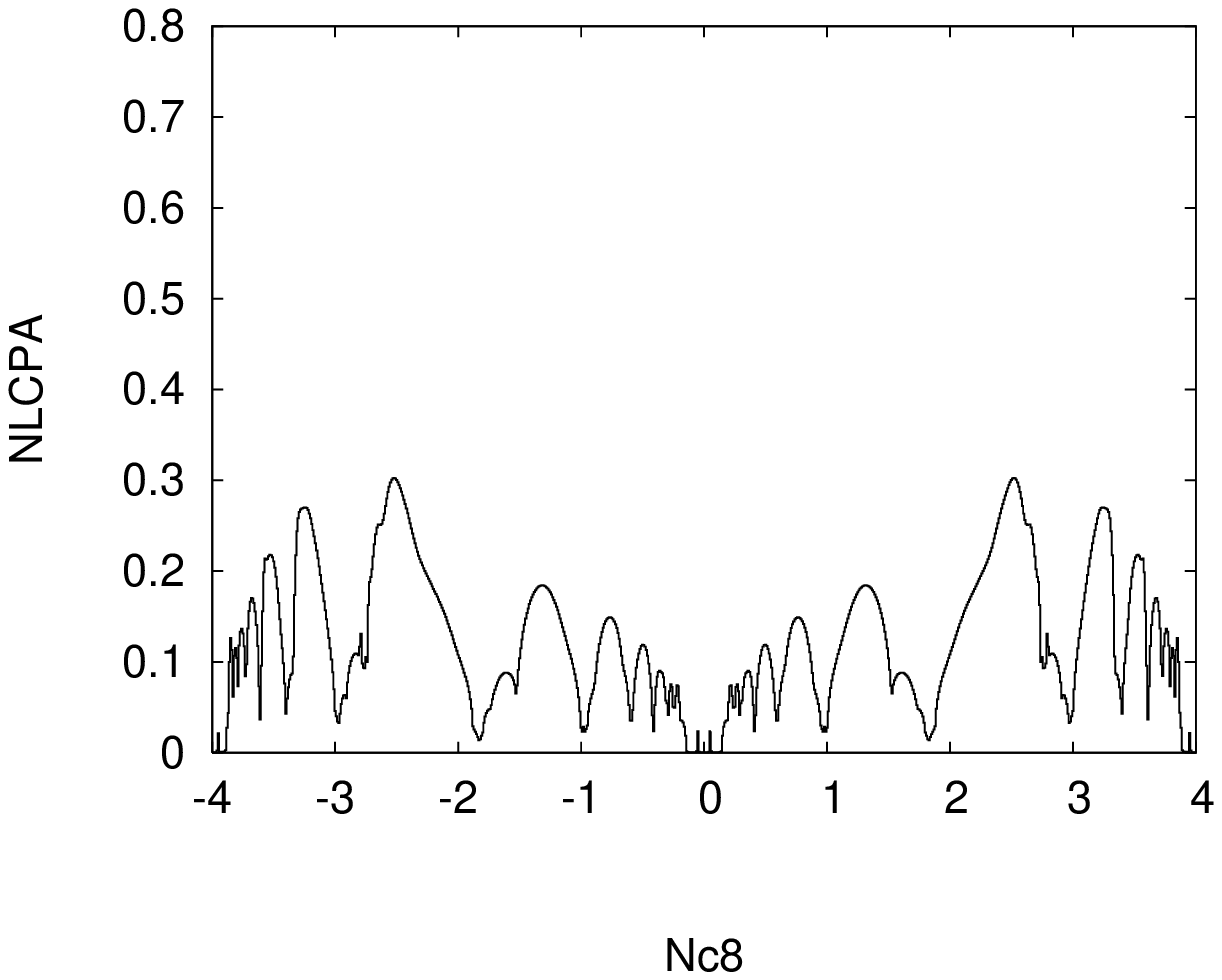}}
 & \psfrag{Nc12}[B1][B1][2.5][0]{$N_C=12$}\scalebox{0.33}{\includegraphics{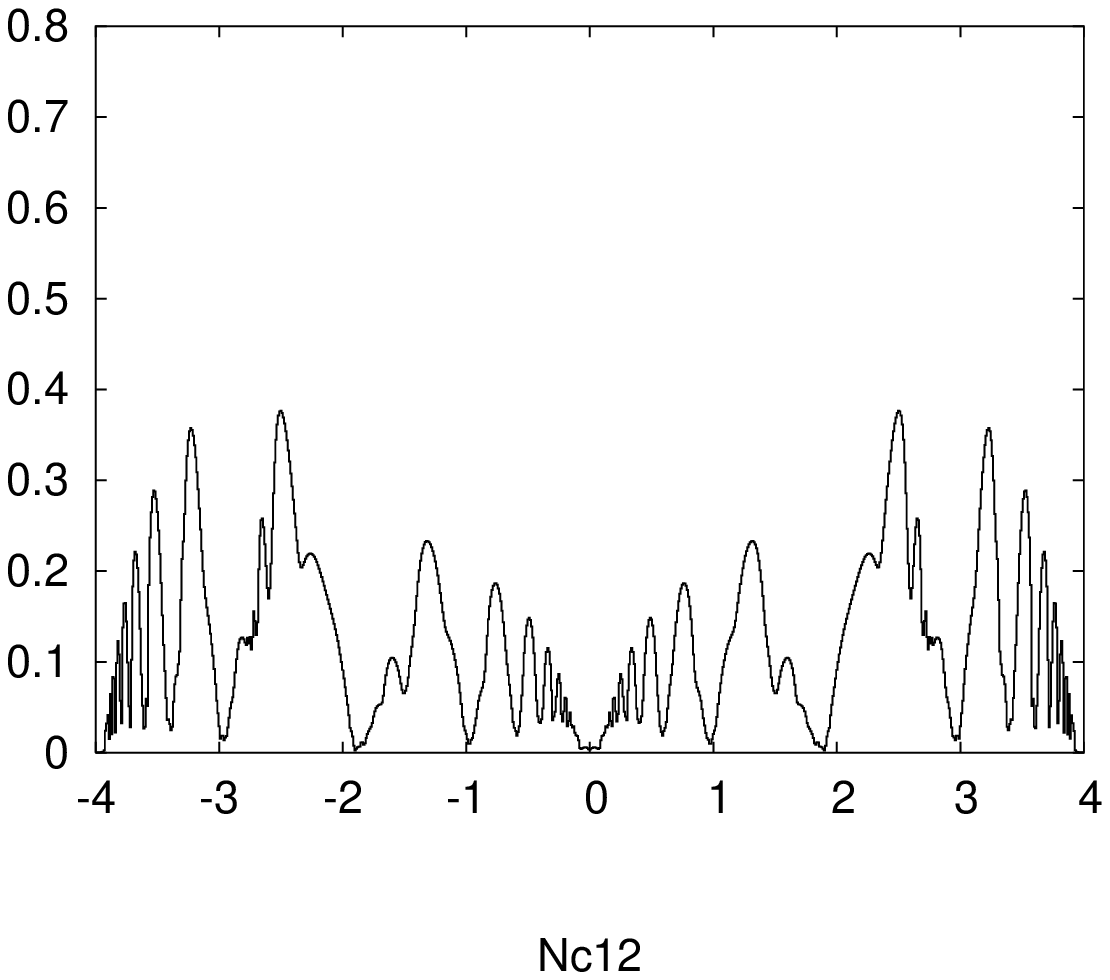}}   
 & \psfrag{Nc16}[B1][B1][2.5][0]{$N_C=16$}\scalebox{0.33}{\includegraphics{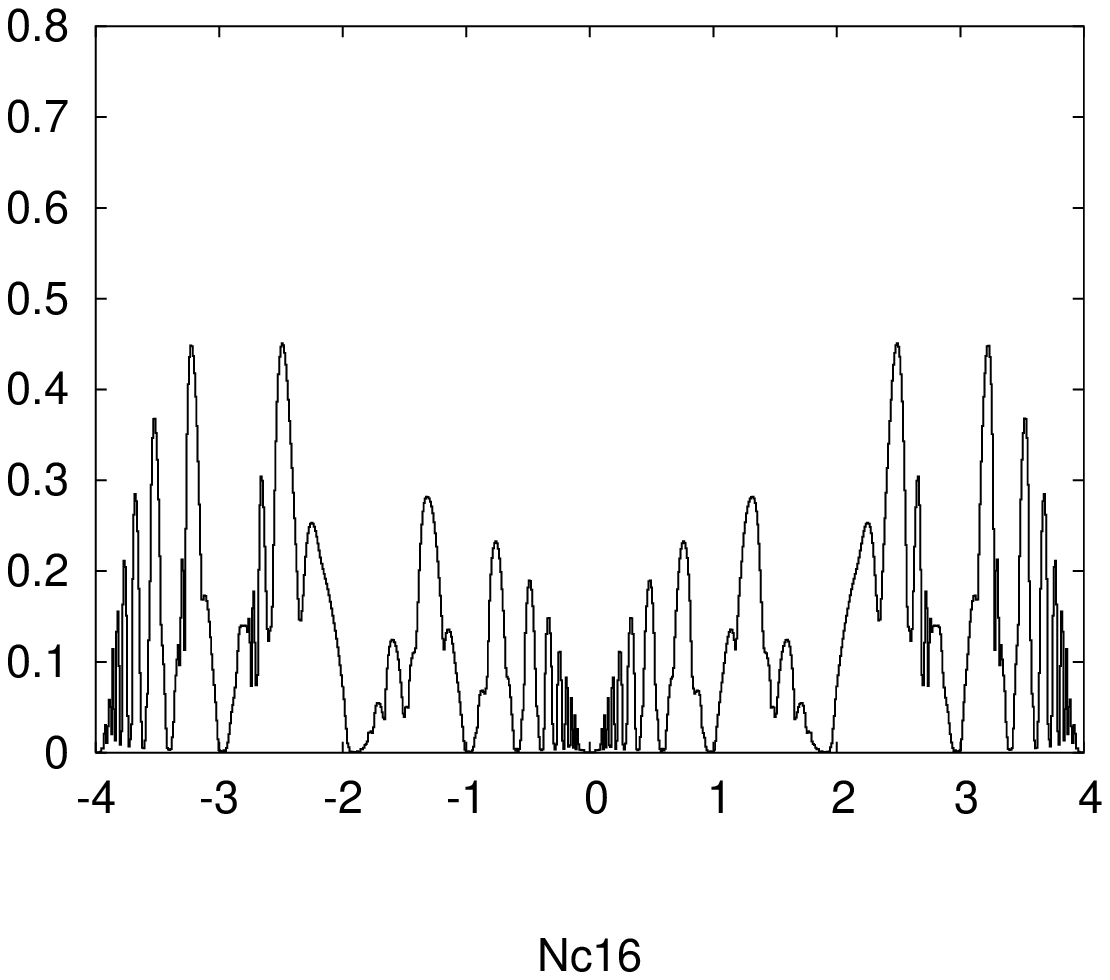}} 
 \end{tabular}         
 \caption{Configurationally-averaged DOS per site as a function of energy (in units of the bandwidth) for the various cluster theories (top 
 to bottom) with cluster sizes $N_c=8,12,16$ (left to right).}\label{cluster81216}
 \end{center}
\end{figure*}

The ability of the various cluster theories to improve upon the CPA result is now examined. All results here have been displayed using
centred histograms at the same energy resolution as the exact result shown in figure~\ref{pureexactcpa}(c), and all energies have a $10^{-3}$
imaginary part.

Figures~\ref{cluster246},\ref{cluster81216} show the DOS for the random~\footnote{An investigation of the ability of these cluster theories
to deal with the effects of chemical short-range order is deferred to another publication.} $A_{50}B_{50}$ alloy of figure~\ref{pureexactcpa}
using the various cluster theories (top to bottom) as a function of cluster size (left to right). For the ECM calculations shown in the first
row, the configurational average has been taken over the $2^{N_c}$ possible cluster configurations and the DOS measured on one of the central
sites. The central site is likely to give a better description of the alloy since it is more influenced by the effects of the disorder than
say a site on the boundary. Similarly, the MCPA calculations on the second row (`MCPA centre') have been measured at one of the central
sites. For the MCPA calculations on the third row (`MCPA average'), the DOS has been measured at each site on the cluster, and then an
average taken over all such measurements. The fourth and fifth rows show NLCPA calculations using the periodic and anti-periodic sets of
cluster momenta $\{\b{K}_n^{P}\}$ and $\{\b{K}_n^{A}\}$ respectively. Finally, the sixth row (`NLCPA average') shows an average of the
periodic and anti-periodic results given in the fourth and fifth rows.

First let us consider the ECM and MCPA central site results shown in the first two rows. As a general observation, it can be seen that all
plots show increasing structure in the DOS as the cluster size increases. This arises from statistical fluctuations in the local environment
of a site (smoothed out by the single-site mean-field CPA) due to specific impurity cluster configurations, and can be identified with
corresponding structure in the exact DOS shown in figure~\ref{pureexactcpa}(c). Furthermore, the gap at $E=0.0$ is systematically filled in
as $N_c$ increases due to larger cluster configurations present which contribute to the DOS at the band edges. Although this occurs more
quickly with the MCPA than the ECM, it is interesting to observe that the ECM results and the MCPA central site results are very similar for
all cluster sizes, indicating that self-consistency with respect to the cluster does not have a significantly large effect on the DOS results
for this system (in the absence of chemical SRO). Indeed, the ECM and MCPA central site DOSs become identical for $N_c=12$, and both converge
almost to the exact result at $N_c=16$.~\footnote{In theory, the MCPA has the incorrect boundary conditions as $N_c\rightarrow\infty$ since
the self-energy would become independent of $\b{k}$. In practice, however, the DOS converges quickly to the exact result.} Next, consider the
`MCPA average' results shown in the third row. For $N_c=2$, this is identical to the MCPA central site result by symmetry. As $N_c$
increases, more structure is again present in the DOS, although it is evident that the convergence towards the exact result is much slower.
For example, the troughs in the DOS deepen slightly as $N_c$ increases from 8 to 16, but a higher cluster size would be needed for the
troughs to reach the zero axis as seen in the exact result.

Now let us turn our attention to the NLCPA calculations. First consider $N_c=2$. In the periodic (P) result, the peaks in the DOS can be
identified with corresponding peaks present in the ECM and MCPA calculations, but the troughs present at $E=\pm{3}$ and $E=\pm{1.5}$ are not
seen elsewhere. The anti-periodic (AP) result is equivalent to the CPA here due to symmetry. On the other hand, the average over the P and AP
results shown in the sixth row resembles the MCPA result much more closely, with peaks and troughs lying at the same energies. For $N_c=4$,
both the P and AP calculations separately resemble the ECM and MCPA only over certain energy regions. However, the average over both the A
and AP results again closely resembles those of the ECM and MCPA. Thus already, it is clear that P and AP calculations separately are not
sufficient to yield a physical DOS, but the average does yield a physically meaningful DOS with peaks and troughs which can be associated
with correlations in the cluster disorder configurations. The reason behind this is clear from section~\ref{solutions}; the P and AP
calculations separately reproduce correlations arising from differing regions of reciprocal space, and thus a combination of the two
calculations is necessary to reproduce all correlations within the range of the cluster size. The more sophisticated method for combining the
A and AP solutions \cite{Batt1} mentioned in section~\ref{solutions} is likely to give even better results than the straight average
displayed here, for example the `flat' regions in the $N_c=2$ calculation are likely to be smoothed out. Also, observe that as $N_c$
increases further, the A and AP plots appear increasingly similar, becoming almost identical at $N_c=12$. This is expected since
$\{\b{K}_n^{P}\}$ and $\{\b{K}_n^{AP}\}$ become closer together as $N_c$ increases. 

Next, let us consider the convergence properties of the NLCPA. Importantly, it is clear that, provided the periodic and anti-periodic results
are combined, the DOS calculated using the NLCPA converges smoothly and systematically towards the exact result. There are however clear
differences in the convergence properties of the NLCPA compared to the ECM and MCPA. Firstly, the gap centred at $E=0.0$ is filled in more
quickly. Secondly, the convergence for the NLCPA is more systematic in that the maximum height of the DOS increases monotonically as $N_c$
increases ($N_c=20$ and $N_c=24$ calculations further confirm this). On the other hand, this means that large `spikes' present in the ECM and
MCPA DOS are not reproduced for small cluster sizes, although such spikes are seen in the exact result. Indeed, the ECM and MCPA plots show
much more fine structure for a given cluster size, whilst the NLCPA results are much more `smeared out'. Consequently, the ECM and MCPA
central site calculations converge much more quickly to the exact result as $N_c$ increases. However, it is important to appreciate that the
main (and more important) features of the exact result are reproduced fairly quickly by the NLCPA. Also, the ECM and MCPA calculations are
very sensitive to the energy resolution at which they are displayed, unlike the NLCPA. These differences in the convergence properties are
all a consequence of the fact that the NLCPA is based on Nyquist's sampling theorem \cite{Elliot1,Jarrell1}. Finally, observe that both the
MCPA average (third row in figures~\ref{cluster246},\ref{cluster81216}) and the NLCPA have not yet converged to the exact result at $N_c=16$.
Indeed there has been much debate in the literature~\cite{Maier1,Biroli1,Aryanpour1,Biroli2} concerning whether the dynamical cluster
approximation (DCA)~\cite{Hettler1} or the cellular dynamical mean field theory (CDMFT)~\cite{Kotliar1} (the analogues of the NLCPA and MCPA
respectively for strongly-correlated electron systems) converges more quickly, carried out in terms of an average hybridization function
coupling the cluster to the medium. Whilst a convergence comparison of the DOS using the NLCPA and MCPA average is not one of the objectives
of this paper, it would be interesting to carry this out by using larger (and computationally very demanding) cluster sizes. It is clear,
however, that a convergence comparison depends greatly on whether the quantity in question is sensitive to the energy resolution. For
example, the convergence properties are likely to be much more similar for a quantity such as the integrated DOS.

\section{Conclusions}\label{conclude}

A comparison of the formalism of the NLCPA with the ECM and MCPA has been presented. In the ECM and MCPA, the cavity function introduced has
an explicit real-space expansion and so the coupling of an impurity cluster to the surrounding medium is well defined. It is found that this
is not the case for the NLCPA cavity function (for $N_c>1$) as a consequence of the introduction of Born-von Karman boundary conditions. This
implies that the cavity should be viewed as a mathematical construction used to determine the medium, in departure from traditional effective
medium theories. The consequences of introducing anti-periodic Born-von Karman boundary conditions within the NLCPA has also been
investigated. Numerical DOS results for a 1D model have been compared. It can be concluded that, firstly, the periodic and anti-periodic
solutions must be combined when using the NLCPA, particularly for small cluster sizes. Importantly, only by then comparing the DOS with the
exact result does it become evident that the NLCPA does indeed produce physical results and, for each cluster size, shares main features in
common with both the ECM and MCPA. Since the NLCPA is based upon Nyquist's sampling theorem, it is however found that the DOS is far more
`smeared out' for a given cluster size. Whilst the NLCPA DOS does (importantly) converge systematically towards the exact result as the
cluster size increases, it therefore does so less quickly than the ECM and MCPA (measured on the central site and in the absence of chemical 
SRO). On the other hand, such fine structure is less important and in any case would be smeared out in higher dimensions. More importantly, 
the NLCPA preserves the full symmetry of the underlying lattice, yields well-defined site-diagonal and site off-diagonal properties, and is 
computationally tractable for realistic systems.

\ack

The author thanks G.~M.~Batt, A.~Gonis, B.~L.~Gy{\"{o}}rffy, and M.~Jarrell for useful discussions, and D.~K{\"{o}}dderitszch for the exact
histogram results. This work was funded by a research fellowship in theoretical physics from EPSRC~(UK), grant GR/S92212/01.

\end{document}